\documentclass{elsart}
\usepackage{natbib}
\usepackage{epsf}

\newcommand{\gtrsim}{\mathrel{\raisebox{0.4ex}{\hbox{$>$}}\kern-0.75em\raisebox{-0.5ex}{\hbox{$\sim$}}}}
\newcommand{\lesssim}{\mathrel{\raisebox{0.4ex}{\hbox{$<$}}\kern-0.75em\raisebox{-0.5ex}{\hbox{$\sim$}}}}
\newcommand{\psim}{\mathrel{\raisebox{0.4ex}{\hbox{$\propto$}}\kern-0.75em\raisebox{-0.5ex}{\hbox{$\sim$}}}}
\newcommand{\gsim}{\mathrel{\raisebox{0.4ex}{\hbox{$>$}}\kern-0.75em\raisebox{-0.5ex}{\hbox{$\sim$}}}}
\newcommand{\lsim}{\mathrel{\raisebox{0.4ex}{\hbox{$<$}}\kern-0.75em\raisebox{-0.5ex}{\hbox{$\sim$}}}}
\def\beq{\begin{equation}}
\def\eeq{\end{equation}}
\def\ba{\begin{array}}
\def\ea{\end{array}}

\def\apj{Astrophys.~J.~}

\begin{document}
\runauthor{Wick, Dermer and Atoyan}
\begin{frontmatter}
\title{High-Energy Cosmic Rays from Gamma-Ray Bursts}
\author[NRL,NRC]{Stuart D.~Wick,}
\author[NRL]{Charles D.\ Dermer,}
\author[UdM]{and Armen Atoyan}

\address[NRL]{Code 7653, Naval Research Laboratory, 
Washington, DC 20375-5352 U.S.A.}
\address[NRC]{NRL/National Research Council Research Associate}
\address[UdM]{CRM, Universit\'e de Montr\'eal, Montr\'eal, Canada H3C 3J7}

\begin{abstract}
A model is proposed for the origin of cosmic rays (CRs) from $\sim
10^{14}$ eV/nucleon to the highest energies ($\gtrsim 10^{20}$ eV).
GRBs are assumed to inject CR protons and ions into the interstellar
medium of star-forming galaxies---including the Milky Way---with a
power-law spectrum extending to a maximum energy $\sim 10^{20}$ eV.
The CR spectrum near the knee is fit with CRs trapped in the Galactic
halo that were accelerated and injected by an earlier Galactic GRB.
These CRs diffuse in the disk and halo of the Galaxy due to
gyroresonant pitch-angle scattering with MHD turbulence in the
Galaxy's magnetic field.  The preliminary (2001) KASCADE data through
the knee of the CR spectrum are fit by a model with energy-dependent
propagation of CR ions from a single Galactic GRB.  Ultra-high energy
CRs (UHECRs), with energies above the ankle energy at $\gtrsim 3
\times 10^{18}$ eV, are assumed to propagate rectilinearly with their
spectrum modified by photo-pion, photo-pair, and expansion losses.  We
fit the measured UHECR spectrum assuming comoving luminosity densities
of GRB sources consistent with possible star formation rate histories
of the universe.

For power-law CR proton injection spectra with injection number index
$p \gtrsim 2$ and low and high-energy cutoffs, normalization to the
local time- and space-averaged GRB luminosity density implies that if
this model is correct, the nonthermal content in GRB blast waves is
hadronically dominated by a factor $\approx 60$-200, limited in its
upper value by energetic and spectral considerations.  Calculations
show that 100 TeV -- 100 PeV neutrinos could be detected several times
per year from all GRBs with kilometer-scale neutrino detectors such as
IceCube, for GRB blast-wave Doppler factors $\delta \lesssim 200$.
GLAST measurements of $\gamma$-ray components and cutoffs will
constrain the product of the nonthermal baryon loading and radiative
efficiency, limit the Doppler factor, and test this scenario.
\end{abstract}
\end{frontmatter}

\section{Introduction}

In this paper we develop a model for high energy cosmic rays (HECRs;
here defined as $\gtrsim 10^{14}$eV/nucleon CRs), based on the
underlying assumption that CRs are accelerated in the relativistic and
nonrelativistic shocks found in GRBs and their attendant supernovae
(SNe). This model extends the work of Vietri \cite{vie95} and Waxman
\cite{wax95}, who proposed that UHECRs originate from GRBs (see also Ref.\
\cite{mu96}), to include cosmic ray production from SNe and GRB sources
in our Galaxy \cite{der02}. To test the model, KASCADE
\cite{ulr01,ber01,kam01} and HiRes-I and HiRes-II Monocular data
\cite{hires} are fit over the energy range $\approx 2\times 10^{16}$
eV to $\approx 3\times 10^{20}$ eV.  A good fit to the entire data set
is possible if CRs are injected with a power-law spectrum with number
index $p = 2.2$, as could be expected in a scenario where particles
are accelerated by relativistic shocks \cite{bo98,kirk}.
 
CRs that are injected with energies $\lesssim 10^{19}$ eV diffuse
through and escape from their host galaxy.  The sources of high-energy
CRs, namely GRBs, are located in star-forming regions found in the
galaxy's disk.  CR transport in the disk and halo of the Milky Way is
modeled using a time-dependent, spherically-symmetric propagation
model that employs an energy-dependent, spatially-independent
diffusion coefficient.  The random-walk pathlengths are assumed to
arise from gyroresonant, pitch-angle scattering of CRs with a
magnetohydrodynamic (MHD) turbulence spectra that can be decomposed
into two components reflecting different power-law distributions of
turbulence over different wavelength ranges. The spectral break at the
CR knee energy is explained in an impulsive, single-source model for
HECRs if the turbulence spectral index changes from Kraichnan to
Kolmogorov turbulence near the wavenumber resonant with CRs at the
knee of the CR spectrum.  We model the preliminary KASCADE data
reported in 2001 by assuming that all ionic species have the same
injection index, with the compositions of the ions adjusted to fit the
observed CR spectrum near the knee.  Additional hardenings of the
low-energy CR spectrum from a single GRB source of HECRs can result
both from energy-dependent diffusion and from a low-energy cutoff in
the CR injection spectrum. Superposition of the contributions from
many SNe are assumed to accelerate the bulk of the GeV/nuc -- TeV/nuc
CRs, as in the conventional scenario \cite{gs64,hay69}.
 
UHECRs have such large gyroradii and diffusion mean free paths that
they are assumed to escape directly from the halo of the GRB host
galaxy and stream into metagalactic space.  The UHECR energy spectrum
evolves in response to photo-pair and photo-pion losses on the
redshift-dependent cosmic microwave background radiation (CMBR), which
we treat in a continuous energy-loss formalism. UHECRs in intercluster
space also lose energy adiabatically due to cosmic expansion. The
measured UHECR spectrum arises from the contributions of sources
throughout the universe, with an intensity that depends on the local
luminosity density of GRB sources and the evolution of the GRB
luminosity density with redshift (see Refs.\ \cite{nw00} and
\cite{mes02} for recent reviews of UHECRs and GRBs, respectively). The
UHECR spectral model assumes that many sources produce the measured
ultra-high energy and super-GZK ($\geq 10^{20}$ eV) CRs.

Our model for UHECRs from GRBs implies a local time- and
space-averaged CR luminosity density $ \dot \varepsilon_{CR} =
f_{CR}\dot\varepsilon_{GRB,X/\gamma}$ of CRs. The local ($z\ll
1$) luminosity density $\dot \varepsilon_{GRB,X/\gamma}\approx
10^{44}$ ergs Mpc$^{-3}$ yr$^{-1}$ is inferred from BATSE observations
of the hard X-ray/soft $\gamma$-ray (X/$\gamma$) emission from GRBs
\cite{wb99,bd00,der02}.  The value of $\dot\varepsilon_{CR}$ depends
sensitively on the minimum energy $E_{min}$ of CR injection for soft
injection spectral indices $p \gtrsim 2$.  For $p \simeq 2.2,$ if
$E_{min}=1$ GeV then $\gtrsim 700\times$ more energy must be injected
in nonthermal hadrons than is observed as X/$\gamma$ emission from
GRBs. If $E_{min}\approx 100$ TeV, then $\approx 70\times$ more energy
is required. Such large baryon loads would provide a bright cascade
emission signature in GRB spectra at MeV -- GeV energies through
photopion or hadron synchrotron processes \cite{bd98,zm01}.

The large nonthermal baryon load, $f_{CR}\gg 1$, required to fit the
UHECR data assuming that the GRBs comoving luminosity density traces
the star formation rate (SFR) history, implies that GRBs can be much
more luminous neutrino sources than predicted under the standard
assumption that the energy injected into a GRB blast wave in the form
of CRs is equal to the energy inferred from X/$\gamma$ fluence
measurements of GRBs \cite{da03}.  We therefore predict that if this
model is correct, and CRs are accelerated and injected in the form of
soft power-laws with index $p \gtrsim 2$, then IceCube should detect
up to several neutrinos from the brightest GRBs with total X/$\gamma$
radiation fluence at levels $\gg 10^{-4}$ erg cm$^{-2}$. This
prediction holds both in a collapsar scenario when the Lorentz factors
$\Gamma$ of the relativistic outflows are $\lesssim 200$, or if the
GRB takes place in an intense external radiation background for a wide
range of $\Gamma$. Lower limits to $\Gamma$ can be inferred from
$\gamma$-ray transparency arguments applied to observations of GRBs
with the {\it Gamma-ray Large Area Space
Telescope\footnote{http://www-glast.stanford.edu/}}.

We also consider whether this model can explain the AGASA data
\cite{tak98} for the UHECRs. Poor fits are found if GRBs inject soft
CR spectra with $p \gtrsim 2$. However, if GRBs inject hard spectra
with $p \approx 1$, for example, through a second-order relativistic
shock-Fermi process \cite{dh01} or through the converter mechanism
\cite{derishev03}, then the highest-energy AGASA data can be fit, though
the reduced $\chi^2$ of our best fits are not compelling. Because the
injection spectrum is so hard, most of the produced high-energy
neutrinos are too energetic and the flux too weak to be detected with
IceCube, though other telescope arrays, such as the Extreme Universe
Space Observatory (EUSO), could be sensitive to these GZK
neutrinos. To explain the CRs below $\approx 3\times 10^{19}$ eV, an
additional component of CRs would still be required either from GRBs
or another class of sources, and these would make an additional
contribution to high-energy neutrino production.

Section 2 gives a discussion of the CR and photon luminosity density
of GRBs and their event rate.  Our propagation model describing CR
diffusion in the disk and halo of the Galaxy is presented in Section
3, where we fit the KASCADE data between $\approx 0.8$ and 200 PeV
with a single Galactic GRB source $\approx 500$ pc away that took
place around 200,000 yrs ago.  In Section 4, we describe our
calculation of the UHECR flux, including energy losses from cosmic
expansion, and photo-pair and photo-pion production.  We present
minimum $\chi_r^2$ fits to the high-energy KASCADE, HiRes-I and
HiRes-II Monocular data covering the energy range $2\times 10^{16}$~eV
to $3\times 10^{20}$~eV. We also fit the AGASA data from $3\times
10^{19}$~eV to $3\times 10^{20}$~eV for hard CR injection
spectra. Section 5 presents new high-energy neutrino calculations from
hadronically dominated GRBs, and our predictions for km-scale
high-energy neutrino telescopes such as IceCube or a deep underwater,
northern hemisphere array. Discussion of the results and conclusions
are given in Section 6.  Our treatment of the UHECR attenuation and
flux calculation is described in Appendix A.

\section{Gamma Ray Burst Model for High-Energy Cosmic Rays}

The progenitor sources of HECRs are likely to be GRBs, as suggested
by the following observations:

\begin{enumerate}

\item The evidence from KASCADE \cite{kam01} that
the break energies of the different CR ionic species are proportional
to rigidity, and that the mean atomic mass increases with $E$ through
the knee region, admits a propagation solution to understand the
energies of the spectral breaks of different CR ions---given a
galactic source that injects power-law CRs to the highest energies.

\item Composition changes above the ``second knee" at 
$E\approx 10^{17.6}$ eV, and possible GZK features in the UHECR
spectrum \cite{bw03,mbo03}, implies that UHECRs and their sources are
metagalactic.

\item The CR all-particle spectrum  breaks at $\cong 3$ PeV 
by $\approx 0.3$ units and then extends without spectral change, other
than for a possible weak softening above the second knee, to the ankle
at $E\approx 3\times 10^{18}$ eV \cite{som01}.  A single power-law
injection source, modified by acceleration and transport effects,
provides the simplest solutions.

\end{enumerate}

A comprehensive model of HECRs therefore seems achievable if a source
type that injected power-law distributions of relativistic CRs were
found in star-forming $L^*$ galaxies such as the Milky Way, with
injection episodes frequent and energetic enough to power the
HECRs. For reasons reviewed in Refs.\ \cite{der02} and
\cite{der01a}, GRBs offer the most likely solution.


\subsection{Local GRB and Super-GZK Emissivities}

We first address the question of the local ($z \ll 1$) time- and
space-averaged GRB luminosity density (or emissivity)
$\dot\varepsilon_{GRB,X/\gamma}$ (ergs Mpc$^{-3}$ yr$^{-1}$) and the
local luminosity density $\dot\varepsilon_{GZK}$ required to power the
super-GZK ($\geq 10^{20}$ eV) CRs.  Pre-Beppo SAX estimates
\cite{vie95,wax95} concerning the X/$\gamma$ emission from GRBs found
\begin{equation}
\dot \varepsilon_{GRB,X/\gamma}\simeq 10^{44} 
\dot\varepsilon_{44} {\rm ~ergs~Mpc^{-3}~yr}^{-1}\;,
\label{elldot}
\end{equation}
with $\dot\varepsilon_{44} \approx few$. This exceeds the local ($z
\ll 1$) GZK emissivity $\dot \varepsilon_{GZK}\simeq u/t_{\phi\pi}
\simeq 6\times 10^{43} u_{-21}$ ergs Mpc$^{-3}$ yr$^{-1}$ required to
power $> 10^{20}$ eV CRs, where $10^{-21} u_{-21}$ ergs cm$^{-3}$ is
the observed energy density in $> 10^{20}$ eV CRs ($u_{-21}\cong 0.5$
for HiRes and $u_{-21} \cong 2$ for AGASA), and the photo-pion
energy-loss timescale for a $10^{20}$ eV proton is $140$ Mpc/$c$
\cite{sta00}. A better estimate that corrects for the
energy-dependence of $t_{\phi\pi}$ implies $\dot
\varepsilon_{GZK}\simeq u/t_{\phi\pi} \simeq 10^{44} u_{-21}$ ergs
Mpc$^{-3}$ yr$^{-1}$, comparable with the GRB emissivity
$\dot\varepsilon_{GRB,X/\gamma}$, so that the required luminosity
density to power the super-GZK CRs is in coincidence with that
available from GRBs.

Based on the study by B\"ottcher and Dermer
\cite{bd00} of BATSE statistics in the external shock model,
 a value of $\dot\varepsilon_{44} \cong 4$ was derived \cite{der02}
using an analytic fit to the SFR derived from Hubble Deep Field
measurements \cite{Madau}.  This emissivity includes, in addition to
the photon luminosity that is dominated by hard X-ray and $\gamma$-ray
emission, an inefficiency factor for production of photons due to the
radiative regime implied by the fits to the data \cite{bd00}, and a
further correction for dirty and clean fireballs \cite{dcb99}, which
have since been discovered \cite{Heise}.

Removing the inefficiency factor from the estimate of Ref.\
 \cite{der02} gives the local GRB emissivity in the form of nonthermal
 photon radiation to be $\dot\varepsilon_{44} \cong 0.6$. Vietri, de
 Marco, and Guetta \cite{vmg03} argue for a value of
 $\dot\varepsilon_{44} \cong 1.1$. Given the various uncertainties, we
 take the local GRB emissivity in the form of hard X-rays
 and 100 keV -- MeV $\gamma$-rays from GRBs to be
\begin{equation}
\dot\varepsilon_{44} \cong 1\;.
 \label{l44}
\end{equation}

\subsection{Beaming and GRB Rate}

A second issue to be considered is the source rate of GRBs in our
Galaxy. Beaming breaks in the optical light curves of GRBs indicate
that the most apparently luminous GRBs are highly collimated. Frail et
al.\ \cite{fra01} argue that a typical GRB has a beaming factor of
1/500$^{th}$ of the full sky, so that GRBs are in actual fact 500
times more numerous, and 1/500$^{th}$ as energetic on average, as the
measured energy releases imply. This means that most GRBs typically
release $\approx 5\times 10^{50}$ ergs in X/$\gamma$ emission.

The local density of $L^*$ spiral galaxies, of which the Milky Way is
representative, can be derived from the Schechter luminosity function,
and is $\approx 1/$(200--500 Mpc$^3$) \cite{wij98,der02}.  The BATSE
observations imply $\sim 2$ GRBs/day over the full sky
\cite{ban02}. Due to beaming, this rate is increased by a factor of
$500 f_{500}$, where $f_{500} \sim 1$. Given that the volume of the
universe is $\sim 4\pi(4000$ Mpc)$^3/3$, this implies a rate per $L^*$
galaxy of $$
\rho_{L^*}\approx {(200-500)~{\rm Mpc}^3/L^*\over {4\pi\over 3}
(4000~{\rm Mpc})^3}\;{700\over {\rm yr}}
\times 500 f_{500}\times SFR\times K_{FT}$$
$$\approx (200-500)~{{\rm Mpc}^3\over L^*} \;\times 0.33\, 
{\rm {isotropic~events \over Gpc^{3}~yr}}
\times 500 f_{500} \times ({SFR\over 1/8})\times K_{FT} 
$$
\begin{equation}
\;\;\;\;\;\;\;\approx (1-3)\times 10^{-4}\;({SFR\over 1/8})
\times ({ K_{FT}\over 3})\times f_{500} \;\;L^{* -1}{\rm ~yr}^{-1}\;.
\label{rhoL*}
\end{equation}
The factor $SFR$ corrects for the star-formation activity at the
present epoch [$SFR(z=0)\cong (1/8)SFR(z=1)$] (see Section 4), and the
factor $K_{FT}$ accounts for dirty and clean fireball transients that
are not detected as GRBs \cite{bd00}. This estimate is in
agreement with the result of Ref.\ \cite{vmg03} that excludes short
GRBs.
Thus a GRB occurs about once every 3--10 millennia throughout
the Milky Way, or at about 10\% of the rate of Type Ib/c
SNe.

We write $\dot\varepsilon_{CR} = f_{CR}\dot\varepsilon_{GRB,X/\gamma}
$ for the local luminosity density of CRs injected by GRBs, where
$f_{CR}$ is the nonthermal baryon-loading factor by which the
emissivity injected by GRBs in the form of hadronic CRs exceeds the
emissivity inferred from direct observations of the X/$\gamma$
emission from GRBs.  In the following, we take $\dot\varepsilon_{44} =
1$ (eq.\ \ref{l44}). From eqs.\ (\ref{rhoL*}) and (\ref{elldot}), we
find that the average apparent isotropic energy release per GRB is
$\cong 10^{53} ( f_{CR}+\dot\varepsilon_{44})/(K_{FT}/3)$ ergs, and
the smaller actual mean energy release per GRB due to beaming and the
consequent larger number of GRB sources is $\cong 2\times 10^{50} (
f_{CR} + \dot\varepsilon_{44})/[f_{500}(K_{FT}/3)$ ergs.

For our canonical model GRB used to fit CR data near the knee, we
take $E_{CR} = 10^{52}$ ergs, corresponding to $ f_{CR} = 50$. A value
of $ f_{CR} \gtrsim 50$ is implied by the data fits if this model for
HECRs is correct (Section 5).  Hence, in a unified model for HECRs
from GRBs, the GRB blast wave must be strongly baryon-loaded with $
f_{CR} \gtrsim 50$, at least during the prompt phase of the GRB when
the acceleration is most rapid.

\section{Galactic Cosmic Rays from GRBs}

\label{sec:gal-halo}

Observations indicate that GRB sources are located in the disks and
star-forming regions of galaxies undergoing active star formation,
such as the Milky Way.  The relativistic ejecta in the GRB explosions
accelerate and inject CRs into the ISM of the GRB host galaxy in the
form of a power law to the highest energies, though possibly with a
low-energy cutoff to the accelerated proton spectrum.  In relativistic
blast waves, this cutoff is expected at $\sim \Gamma^2$ GeV energies,
which could easily reach TeV -- PeV energies for typical GRB blast
waves with $\Gamma \sim 10^2$ -- 300.  UHECR acceleration and
injection probably occurs during the $\gamma$-ray luminous phase of a
GRB, which is on the order of minutes to hours. Acceleration and
injection of lower energy CRs might operate on the Sedov time scale,
which could exceed thousands of years. These acceleration times are
still short compared to the times for particles with the corresponding
energies to diffuse a distance comparable to the disk scale height, as
can be shown from the diffusion properties of the Galaxy derived
below. A GRB source can therefore be treated as an impulsive source of
CR injection.

In this section we present a simplified propagation model for CRs in
the disk and halo of the Galaxy.

\subsection{Diffusion Mean Free Path}

We assume that the Galaxy's disk and halo magnetic field
consists of a large-scale field of mean strength $B = B_{\mu{\rm G}}$ $\mu$G 
on which is superposed a spectrum of MHD turbulence. 
The diffusion mean-free-path $\lambda$ for CRs diffusing
through this field is a function of the Larmor radius
\begin{equation}
r_{\rm L} =  {A m_pc^2 \beta\gamma \over Z eB}\sin\alpha \cong {E\over Q B} 
\cong 3.12\times 10^{12}  {A \gamma\over Z B_{\mu{\rm G}}}\;{\rm cm} 
\cong {A \gamma_6\over Z B_{\mu{\rm G}}}\;{\rm pc} \;,
\label{Larmor}
\end{equation}
where $\gamma= 10^{6}\gamma_6$ is the Lorentz factor of the
relativistic CR proton or ion with atomic mass $A$, $\beta =
\sqrt{1-\gamma^{-2}}$, and the pitch angle $\alpha$ is set equal to
$\pi/2$.  We assume isotropic turbulence, though it is straightforward
to generalize the treatment for particle-scattering properties that
differ in directions parallel and transverse to the plane of the disk.

We consider a very simplified approach to CR transport where particles
diffuse via pitch-angle scattering with resonant MHD turbulence; see,
e.g., Refs.\ \cite{bs87,dml96,cb98,bla00}, and Ref.\ \cite{sch02} for a
detailed treatment.  Let $w(k)dk$ represent the differential energy
density of isotropic MHD turbulence with wavenumbers between $k$ and
$k + dk$. First consider the case where the MHD turbulence spectrum
is described in the inertial range by a single power-law function
$kw(k) = w_0(k/k_1)^{1-q}$ $H(k; k_1,k_2)$, where $H(x;a,b) =1$ when
$a\leq x \leq b$ and $H(x;a,b) =0$ otherwise, and $q$ is the spectral
index of the wave spectrum.  The index $q = 5/3$ for a Kolmogorov
spectrum of turbulence, $q = 3/2$ for a Kraichnan spectrum of
turbulence, and $w_{tot}=w_0/(q-1)$ is the
total energy density of MHD turbulence when $q > 1$ and $k_2 \gg k_1$.
Note that this description of the wave turbulence is an extreme
oversimplification, insofar as we do not consider different wave modes
and helicities, or distinguish between forward- and
backward-propagating waves.

The ratio of the MHD wave energy $w_{tot}$ to the energy density $U_B
\equiv B^2/8\pi$ of the large-scale component of the Galactic
magnetic field is denoted by $\xi_1$, so that $kw(k) \cong (q-1)\xi_1
U_B (k/k_{1})^{1-q}$ for $k_{1} \leq k \leq k_{2}$. The value of
$k_{1}$ corresponds to the inverse of the largest size scale on which
turbulence is injected.

We make the ansatz that the diffusion coefficient of a CR with Larmor
radius $r_{\rm L}$ is inversely proportional to the the energy density
in gyroresonant waves with wavenumber $\bar k \sim r_{\rm L}^{-1}$
\cite{dru83}. We therefore have $\lambda = r_{\rm L}U_B /\bar k w(\bar k) =
k_1^{1-q} r_{\rm L}^{2-q}/\xi_1(q-1)= b_1^{q-1} r_{\rm L}^{2-q}/\xi_1
(q-1)\propto \gamma^{2-q}$, giving
\begin{equation}
\lambda({\rm pc}) = \;  {b_{pc}^{q-1}\over (q-1) \xi_1 } 
({A\gamma_6\over ZB_{\mu{\rm G}}})^{2-q}  = \;\cases{ 
{3b_{pc}^{2/3}\over 2\xi_1} \; 
\;({A\gamma_6\over Z B_{\mu{\rm G}}})^{1/3}\; ,& for $q = 5/3$ \cr\cr 
{2 b_{pc}^{1/2}\over \xi_1} \;({A\gamma_6\over Z 
B_{\mu{\rm G}}})^{1/2}\; ,& for $q = 3/2$ \cr}
\; ,\; \;
\label{lambda}
\end{equation}
where $k_1 = 1/b_1 = b_{pc}^{-1}$ pc$^{-1}$. 

Equation (\ref{lambda}) gives the mean-free path $\lambda$ of
relativistic ($\gamma \gg 1$) CRs that diffuse due to gyroresonant
interactions with large-wavenumber ($k > k_1$) turbulence.  When
$r_{\rm L}^{-1} < k_1$, or $r_{\rm L} > b_{pc}$ pc, CRs gyroresonate
with small-wavenumber ($k < k_1$) turbulence. If the knee in the CR
all-particle spectrum at $E_K\cong 3$ PeV is due to propagation
effects, then $b_{pc} \cong 1$ when $B_{\mu{\rm G}} = 3$ for the
dominant CR proton species.  The break energies in the spectra of
other CR ions are also given by the relation $r_{\rm L} = k_1^{-1} $,
so that the knee energies for the different CR ions occur at energies
\begin{equation}
E_Z ({\rm PeV})\cong A\gamma_6 \cong  Z B_{\mu{\rm G}}b_{pc}\;
 .\; \;
\label{EK}
\end{equation}
The fit to the data described below uses $b_{pc} = 1.6$ and
$B_{\mu{\rm G}} = 3$.  If the value of $\xi_1$, corresponding to the
ratio of the energy density $k \gtrsim k_1$ turbulence to $U_B$ is a
few percent, then $\lambda \simeq h_d$ for $\approx 3$ PeV CR protons,
where $h_d $ is the scale height of molecular gas or massive stars in
the disk of the Milky Way.  A break in the all-particle spectrum due
to a change in the propagation mode occurs at $\approx 3$ PeV, where
$\lambda \approx h_d$, with a pattern of break energies $E_{br}\propto
Z$ for the ionic constituents of the CR all-particle spectrum.
 
Although a CR proton with energy $E_K$ has $r_{\rm L} \sim$ pc $\ll
h_d$, its mean-free-path $\lambda(E_K) \sim h_d \gg r_{\rm L}(E_K)$
because pitch-angle scattering results only from gyroresonant
interactions with MHD waves that constitute a small fraction of the
total magnetic field energy density. CRs injected with $E\gg E_K$ will
diffuse through the Galaxy's disk and halo through gyroresonant
pitch-angle scattering with MHD turbulence carried by waves with
wavenumbers $k \ll k_1$ that gyroresonate with particles with $E \gg
E_K$.  This suggests that we consider a more general wave spectrum
over a larger range of $k$ given by the expression
\begin{equation}
kw(k) = A_0[f_H\;({k\over k_1})^{1-q_H} H(k; k_0, k_1) + 
({k\over k_1})^{1-q} H(k; k_1, k_2)]\;,
\label{kwk}
\end{equation}
where $q_H$ is the index of the small wavenumber MHD turbulence, and
$f_H$ normalizes the energy densities of the small ($k < k_1$) and
large ($k > k_1$) wavenumber turbulence spectra at $k = k_1$.  The
normalization condition $\xi U_B = \int_0^\infty dk \,w(k)$ implies
$A_0 = \xi U_B/ g$, from which the normalization coefficient $g$ can
be easily derived.

\begin{figure}[c]
\centerline{\hbox{
\epsfxsize=400pt \epsfbox{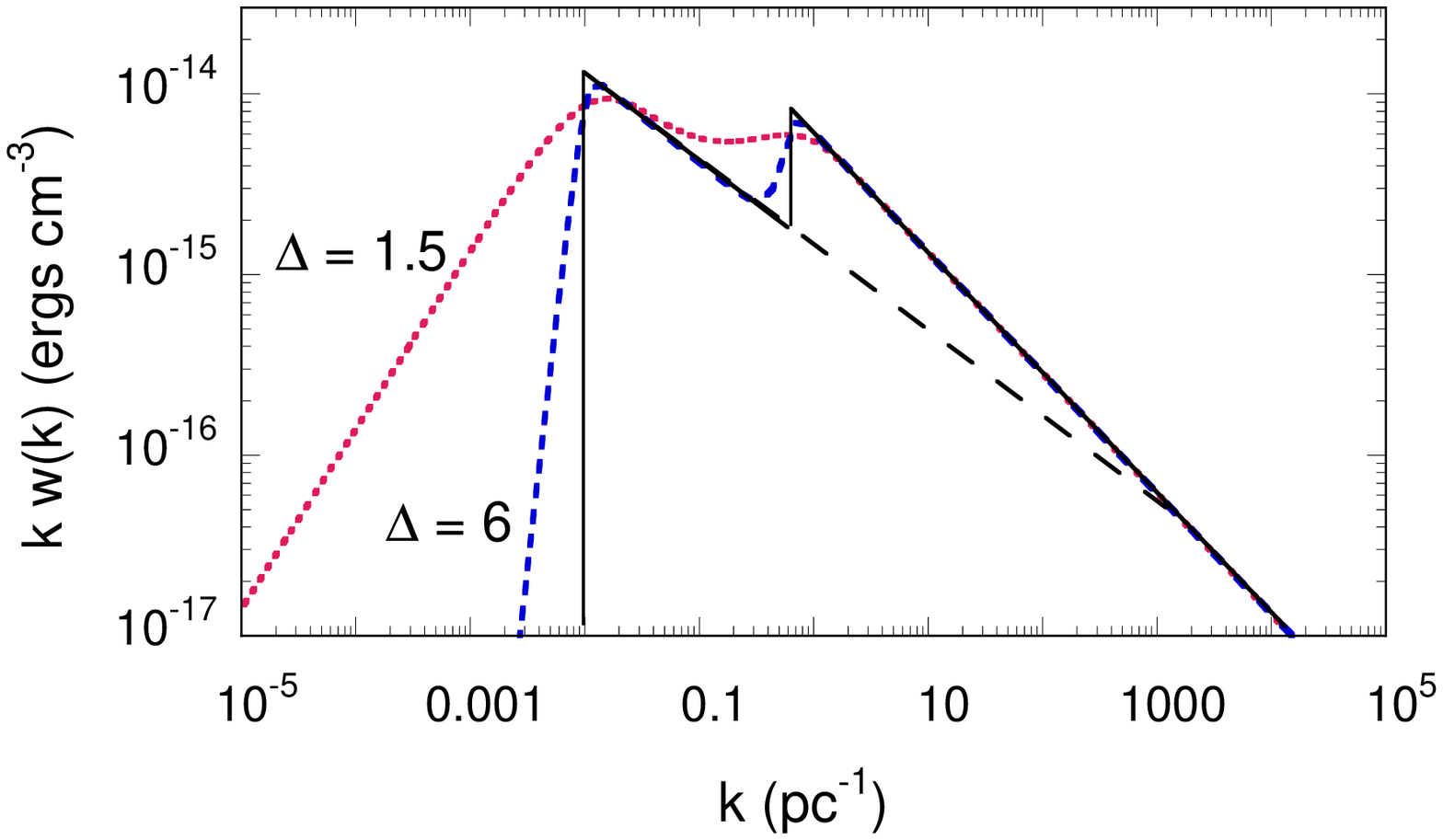}
}}
\caption{ Wave turbulence spectrum used to model 
CR propagation in the Galaxy. The mean magnetic field is assumed to be
$3~ \mu$G.  An idealized model is shown by the solid lines and, after
smoothing, by the dotted and short-dashed curves for the smoothing
parameter $\Delta = 1.5$ and 6, respectively. The parameters for the
model used in the fits are $\xi = 0.1$, $f_H = 0.2$, $k_0=1/100$
pc$^{-1}$, $k_1 = 1/1.6$ pc$^{-1}$, and $\Delta = 1.5$. The extension
of the small-wavenumber ($k < k_1$) turbulence spectrum intersects the
large-wavenumber ($k > k_1$) spectrum at $k \approx 1000$ pc$^{-1}$,
as shown by the long-dashed line.  }
\label{fig:kwk}
\end{figure}

The wave spectrum from eq.~(\ref{kwk}) is plotted in
Fig.~\ref{fig:kwk} by the solid lines.  The parameters are $k_0 =
1/100$ pc$^{-1}$, $k_1 = 1/1.6$ pc$^{-1}$, $f_H = 0.2$, $\xi = 0.1$,
$q_H = 3/2$, $q= 5/3$, and $B = 3\mu$G. Sixty-five percent of the
turbulence energy density is in the small wavenumber component, so
that $\xi_1 = 0.035$. The sharp edges in this wave spectrum introduce
unphysically sharp features in the spectra of particles for this
simple model of transport.  A more physically plausible turbulence
spectrum that retains the features of eq.~(\ref{kwk}) but smooths the
particle distribution function, as would be physically expected from
3-wave interactions and turbulence-energy diffusion and cascading
\cite{sg69}, is given by
\begin{equation}
kw(k) = f_H A_0({k\over k_1})^{1-q_H}\Lambda_
\Delta({k\over k_1})[1-\Lambda_\Delta({k\over k_0})]  
+ A_0 ({k\over k_1})^{1-q}[1-\Lambda_\Delta({k\over k_1})] ,
\label{kwknew}
\end{equation}
where the smoothing function $\Lambda_\Delta(x) = \exp(-x^{-\Delta})$,
and we let $k_2\rightarrow \infty$.  Eq.(\ref{kwknew}) reduces to eq.\
(\ref{kwk}) in the limit $\Delta \gg 1$.  Fig.~\ref{fig:kwk} shows a
plot of eq.\ (\ref{kwknew}) with $\Delta = 1.5$ and, for comparison,
$\Delta = 6$.  The subsequent fits to the data use the MHD wave
spectrum, eq.\ (\ref{kwknew}), with $\Delta = 1.5$.

\begin{figure}[c]
\centerline{\hbox{
\epsfxsize=400pt \epsfbox{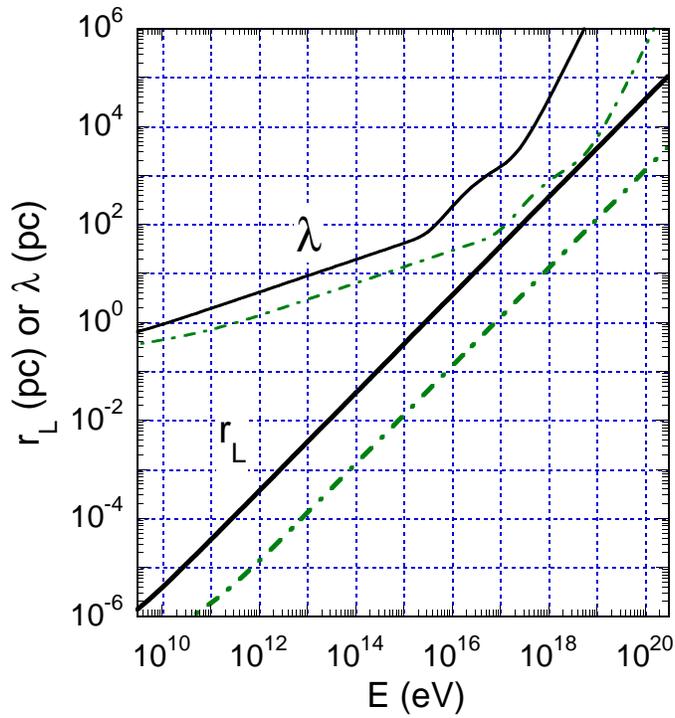}
}}
\caption{  Larmor radius 
$r_{\rm L}$ and mean-free-path $\lambda$ of CR protons (solid curves)
 and Fe nuclei ($Z = 26, A = 56$; dot-dashed curves) with total energy
 $E$ in a magnetic field with mean strength of $3 ~\mu$G.  The wave
 turbulence spectrum given by Fig.~\ref{fig:kwk} with $\Delta = 1.5$
 is used to calculate $\lambda$.  }
\label{fig:lambda}
\end{figure}

The ansatz $\lambda = r_{\rm L}[U_B/\bar k w(\bar k)]$ with $\bar k
\rightarrow r_{\rm L}^{-1}$ therefore implies
\begin{equation}
\lambda \cong ({g \over \xi})\; {r_{\rm L}\over 
f_H (r_{\rm L}k_1)^{q_H-1}\Lambda_\Delta({r_{\rm L}k_1})
[1-\Lambda_\Delta({r_{\rm L}k_0})]  +  (r_{
\rm L}k_1)^{q-1}[1-\Lambda_\Delta({r_{\rm L}k_1})]}.\;
\label{lambdanew}
\end{equation}
In Fig.~\ref{fig:lambda}, we plot $\lambda$ and $r_{\rm L}$ as a
function of total energy $E$ for CR protons and Fe nuclei. The
turbulence spectrum is given by Fig.~\ref{fig:kwk} with $\Delta =
1.5$.  Note that $\lambda \cong 10$ kpc when the CR proton energy
$E\approx 4\times 10^{17}$ eV.  This energy is near the onset of
quasi-rectilinear propagation in a Galactic halo of size $\approx 10$
kpc, and we expect a cutoff in the CR proton flux from a recent GRB in
this energy range whenever we are not in the beam of a GRB jet (which
is nearly always true because of the extreme improbability of such an
occurrence).  The energy-dependent cutoff will be $\propto Z$, as
observed. The actual situation is of course more complicated due to
the magnetic field gradient from the disk to the halo and into
metagalactic space.

\subsection{Three Dimensional Isotropic Diffusion}

We assume that a GRB injects a total energy $E_{CR}= 10^{52} E_{52}$
ergs in the form of cosmic rays. For impulsive injection, the number
injection spectrum of CRs with atomic charge $Z$ and atomic mass $A$
is assumed to be given by $N_{Z,A}(\gamma ) =
c_{Z,A}K(\beta\gamma)^{-p}$ for $(\beta\gamma)_{min} \leq \beta\gamma
\leq (\beta\gamma)_{max}$, and we assume that $p$ is the same for all
($Z,A$). When $(\beta\gamma)_{min}\ll 1$, $(\beta\gamma)_{max}\gg 1$,
and $2 < p < 3$,
\begin{equation}
K \cong [1+{p-2\over 2(3-p)} ]^{-1}\;  
{(p-2)E_{CR} \over m_pc^2  \sum_{Z,A} A c_{Z,A}}\;.
\label{Ka}
\end{equation}
If $1\ll (\beta\gamma)_{min} \cong \gamma_{min} \ll 
(\beta\gamma)_{max}$, $2 < p < 3$, 
\begin{equation}
K \cong   {(p-2)E_{CR} \gamma_{min}^{p-2} \over m_pc^2  
\sum_{Z,A} A c_{Z,A}}\;.
\label{Kb}
\end{equation}

Cosmic rays are assumed to diffuse isotropically in the interstellar
medium, with the initial anisotropy from injection by the GRB jet
quickly washed out. For energy-dependent diffusive propagation, the
spectral number density of relativistic CR ions with charge $Z$ and
mass $A$ measured at a distance $r = 500 r_{500}$ pc away from an
impulsive source of CRs from a GRB that occurred a time $t$ earlier is
\begin{equation}
n_{Z,A}(\gamma; r,t) = {c_{Z,A}K\gamma^{-p}\over \pi^{3/2}
r_{dif}^3}\;\exp[-(r/r_{dif})^2]\;.\;
\label{nZA}
\end{equation}
Here the diffusion radius $r_{dif} \cong
2\sqrt{D(\gamma)t}=2\sqrt{\lambda c t/3}$, and eq.\ (\ref{nZA})
assumes that CRs suffer no significant energy losses during transport
(see Ref.\ \cite{aav95} for the more general case).

Inspection of eq.\ (\ref{nZA}) shows that when the observer is within
the diffusion radius, that is, when $r \ll r_{dif}$, then
\begin{equation}
n_{Z,A}(\gamma; r,t) \propto {\gamma^{-p}\over r_{dif}^{3}}
\propto {\gamma^{-p -{3\over 2}(2-q)}\over t^{3/2}}
\;\propto\;t^{-3/2}\times\;\cases{ 
\gamma^{-p-{1\over 2}}\; ,& for $q = 5/3$ \cr\cr 
\gamma^{-p-{3\over 4}}\; ,& for $q = 3/2$ \cr}\;.\;\;
\label{napprox}
\end{equation}
The measured spectrum from a burstlike source is therefore steepened
by ${3\over 2}(2-q)$ units, where the diffusion coefficient
$D\propto\lambda \propto \gamma^{2-q}$ \cite{aav95}. (By comparison,
the spectral index is steepened by $(2-q)$ units for continuous
injection sources of CRs.)  Eq.\ (\ref{napprox}) shows that
energy-dependent diffusion due to pitch-angle scattering with a
Kolmogorov and Kraichnan spectrum of MHD turbulence steepens the
injection spectrum of an impulsive source by 0.5 and 0.75 units,
respectively. If the disk and halo magnetic fields of the Galaxy
support a two-component MHD turbulence spectrum with indices $q = 5/3$
and $q_H = 3/2$, then an injection spectrum with $p = 2.2$ will be
steepened to a measured spectrum $n_{Z,A}(\gamma; r,t)\propto
\gamma^{-s}$, with $s = 2.7$ at $E\ll E_K$ and $s = 2.95$ at $E \gg
E_K$. Because these indices are similar to the measured CR indices
below and above the knee energy, we adopt this simplified model for CR
transport.  Speculations concerning the origin of such a turbulence
spectrum are deferred to Section 7.

\begin{figure}[c]
\centerline{\hbox{
\epsfxsize=400pt \epsfbox{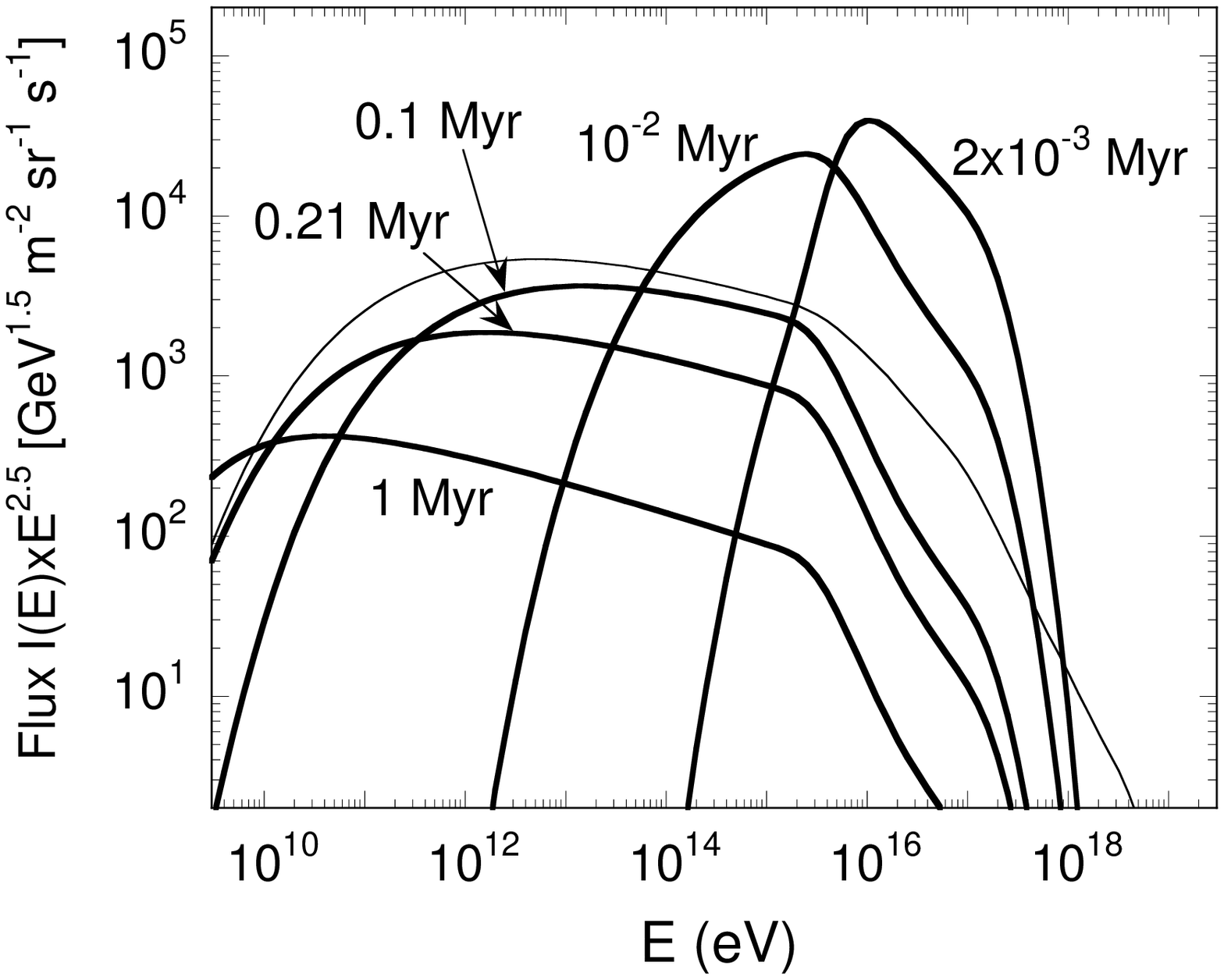}
}}
\caption{Temporal evolution of the fluxes of CR protons 
(heavy curves) measured by an observer at Earth from a GRB that took
place 500 pc away at earlier times given by the labels on the curves.
The GRB was assumed to inject $10^{52}$ ergs total energy in cosmic
rays with an ionic composition given by Table 1 and a low-energy
cutoff at $(\beta\gamma)_{min} = 0.01$ ($\approx 50$ keV).  Shown by
the thin solid curve is the all-particle flux at time $t = 0.21$
Myr used to fit the KASCADE data. Note that the fluxes have been
multiplied by $E^{2.5}$, where $E$ is the CR particle kinetic energy.
}
\label{fig:evolve}
\end{figure}

Fig.~\ref{fig:evolve} shows the CR proton flux measured at Earth at
different times after a GRB explosion located 500 pc from Earth.  The
low-energy cutoff energy is given by $(\beta\gamma)_{min} \ll 1$, so
that most of the energy is deposited by $\approx 1 $ GeV/nuc CRs.
When $(\beta\gamma)_{min} = (\beta\gamma)_{min,1}\gg 1$, the CR
intensity is increased at energies above the low-energy cutoff by the
factor $[(\beta\gamma)_{min,1}]^{p-2}$.

The flux of relativistic CRs is calculated from the expression
\begin{equation}
I_{Z,A}(E) ({\rm m^2~s~sr~GeV)^{-1}} = {10^4 c \over 
4\pi E({\rm GeV})}\;\gamma n_{Z,A}(\gamma;r,t)\;,
\label{IE}
\end{equation}
where $E = \gamma Am_pc^2$ and $n_{Z,A}$ is in cgs units.  Propagation
effects arising from the combined disk and halo MHD turbulence
spectrum produce a break in the spectrum at $E \approx 3$ PeV.  The
index of the CR number fluxes above the knee is $\approx 3.0$, for the
reasons given above, until propagation effects at the ``second knee"
at $E \approx 4\times 10^{17}$ eV start to soften the spectrum so that
the metagalactic component begins to make a dominant contribution.
 
The observer is within the diffusion radius when $r_{dif}(\gamma)\cong
2\sqrt{D(\gamma)t}\gg r$, or when $ct \gtrsim r^2/\lambda $. If we are
considering CRs with energies below the knee, then these particles
scatter with the large wavenumber $k > k_1$, $q = 5/3$ turbulence, so
that $\lambda \cong g k_1^{1-q}r_{\rm L}^{2-q}/\xi$. We write $ct =
3.07\times 10^{5} t_{Myr}$ pc, where the observer is irradiated by a
CR flux from a GRB that took place $t_{Myr}$ Myr ago.  At early times
following the GRB, the low-energy portion of the measured CR spectrum
from a GRB source is exponentially attenuated due to the slower
diffusion of lower energy CRs. The observer will see the maximum flux
from CRs with energies $E = Am_pc^2 \gamma$ when
\begin{equation}
t_{Myr} \approx  0.6({\xi\over g}) r_{500}^2\;
{1\over b_{pc}^{2/3}} ({A\gamma_6\over ZB_{\mu{\rm G}}})^{-1/3}
\approx  0.014 r_{500}^2 ({A\gamma_6\over ZB_{\mu{\rm G}}})^{-1/3}\;\;.
\;\;
\label{tMyr}
\end{equation}
This expression uses the parameters for the model fits discussed
below, with $\xi = 0.1$ and $ g \cong 4.3$, thus showing that the
maximum flux of CR protons near the knee energy reaches an observer
500 pc away $\approx 14000$ years after the GRB, in accord with
Fig.~\ref{fig:evolve}.  Eq.~(\ref{tMyr}) shows that the low-energy
cutoff $\gamma_{cutoff}$ of the CR proton flux in
Fig.~\ref{fig:evolve} evolves according to the relation
$\gamma_{cutoff}\propto r^{1/(1-q/2)}t^{-1/(2-q)}$.  Note that eq.\
(\ref{tMyr}) is only valid when $t> r/c$.

Fits to the CR data near the knee are compatible with a 500 pc distant
GRB releasing $10^{52}$ ergs in cosmic rays if the GRB took place
$\approx 10^5$ yrs ago.  This implies that the 0.1 -- 100 PeV CR
fluxes were up to 100 times brighter one or two hundred thousand years
ago than they are today, though no test of this implication suggests
itself.

\subsection{Fits to the KASCADE Data}

\begin{figure}[c]
\centerline{\hbox{
\epsfxsize=400pt \epsfbox{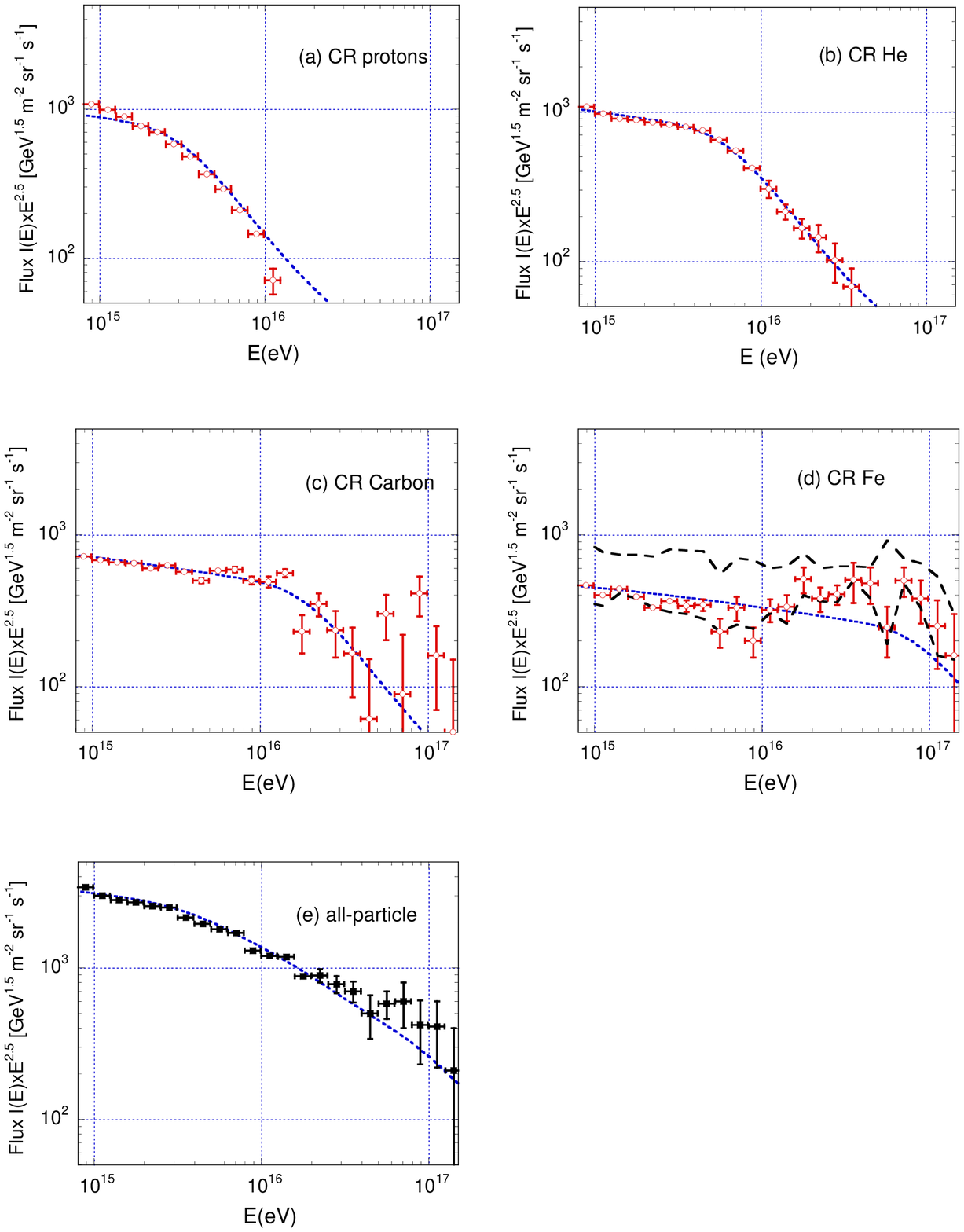}
}}
\caption{  Data points show preliminary KASCADE measurements 
of the CR proton (panel a), He (panel b), Carbon (panel c), Fe (panel
 d), and the all-particle spectrum (panel e), along with model fits
 (dotted curves) to the CR ionic fluxes.  In the model, a GRB that
 occurred $2.1\times 10^5$ years ago and at a distance of 500 pc
 injects $10^{52}$ ergs in CRs, with ionic compositions given in Table
 1.  The CRs isotropically diffuse via pitch-angle scattering with an
 energy-dependent mean-free-path $\lambda$ in an MHD turbulence field
 given by Fig.~\ref{fig:kwk}.  }
\label{fig:KASCADE}
\end{figure}

We now apply this model to fit the preliminary KASCADE data announced
in 2001 \cite{ulr01,ber01,kam01}.  The points in
Fig.~\ref{fig:KASCADE} show KASCADE data and error bars for CR
protons, He, Carbon, Fe, and the all-particle spectrum in panels (a)
-- (e), respectively.  Fig.~\ref{fig:KASCADE}(d) also shows the
reported 1$
\sigma$ boundary to the estimated systematic errors for CR Fe. 

The fits to the KASCADE data shown in Fig.~\ref{fig:KASCADE} use the
propagation model previously described, with a GRB source of CRs at a
distance $r = 500$ pc that exploded 210,000 yrs ago.  The change in
the diffusion and propagation properties leads to a break in the CR
particle spectrum at break energies $\propto Z$ (eqs.~[\ref{lambda}]
and [\ref{EK}]).  Because of the preliminary nature of the data and
the potentially significant systematic errors that could still remain
in the early analyses, we did not perform a rigorous fit to the data,
but instead adjusted the wavenumber $k_1$ and the compositions of the
different ionic species until a reasonable fit to the data was
obtained. With $B = 3 \mu$G, the best value for the energy-dependent
break was obtained with $k_1 \cong 1/1.6$ pc in the spectrum of
turbulence.
 
The Anders-Grevesse Solar photospheric composition \cite{ag89}, and
the (energy-independent) metallicity enhancements, compared to the
Anders-Grevesse compositions, that were used to fit the data, are
listed in Table 1. The strong enhancements by a factor of 50 and 20
for C and Fe, respectively, may be possible from highly-enriched winds
that could be peculiar to a GRB stellar progenitor, or if there was an
earlier supernova as in the supranova model. Note that O cannot be
strongly enhanced if we are to maintain a good fit to the all-particle
spectrum.  A more detailed discussion about implications from
composition will depend on fits to final analysis of KASCADE data
\cite{rot03}.

\begin{table*}
\caption{Cosmic Ray Abundances Relative to 
Solar Photospheric Abundances \cite{ag89}}

\label{defparagcl} 
\begin{center}
\begin{tabular}{l l l}
\hline 
Ion & $c_{Z,A} $   &Enhancement\\  
\hline 
$^1_1{\rm H}$ & 1.00  & 1.0\\  
$^4_2 {\rm He}$ &  0.098 &    1.6\\ 
$^{12}_6{\rm  C}$ & $3.63\times 10^{-4}$   & 50  \\ 
$^{16}_8 {\rm O}$ & $8.51\times 10^{-4}$ &   $\lesssim 5$ \\ 
$^{56}_{26} {\rm Fe}$ & $4.68\times 10^{-5}$   & 20 \\ 
\hline 
\end{tabular}
\end{center}

\vspace*{.6cm}
\end{table*}

As can be seen from Fig.~\ref{fig:KASCADE}, this model gives a
reasonable fit to the data for the CR ion and all-particle spectra
around the knee.  Thus we argue that propagation effects from CRs
injected and accelerated by a single GRB are responsible for the shape
of the CR spectrum between $\approx 1$ and 100 PeV.

\begin{figure}[c]
\centerline{\hbox{
\epsfxsize=400pt \epsfbox{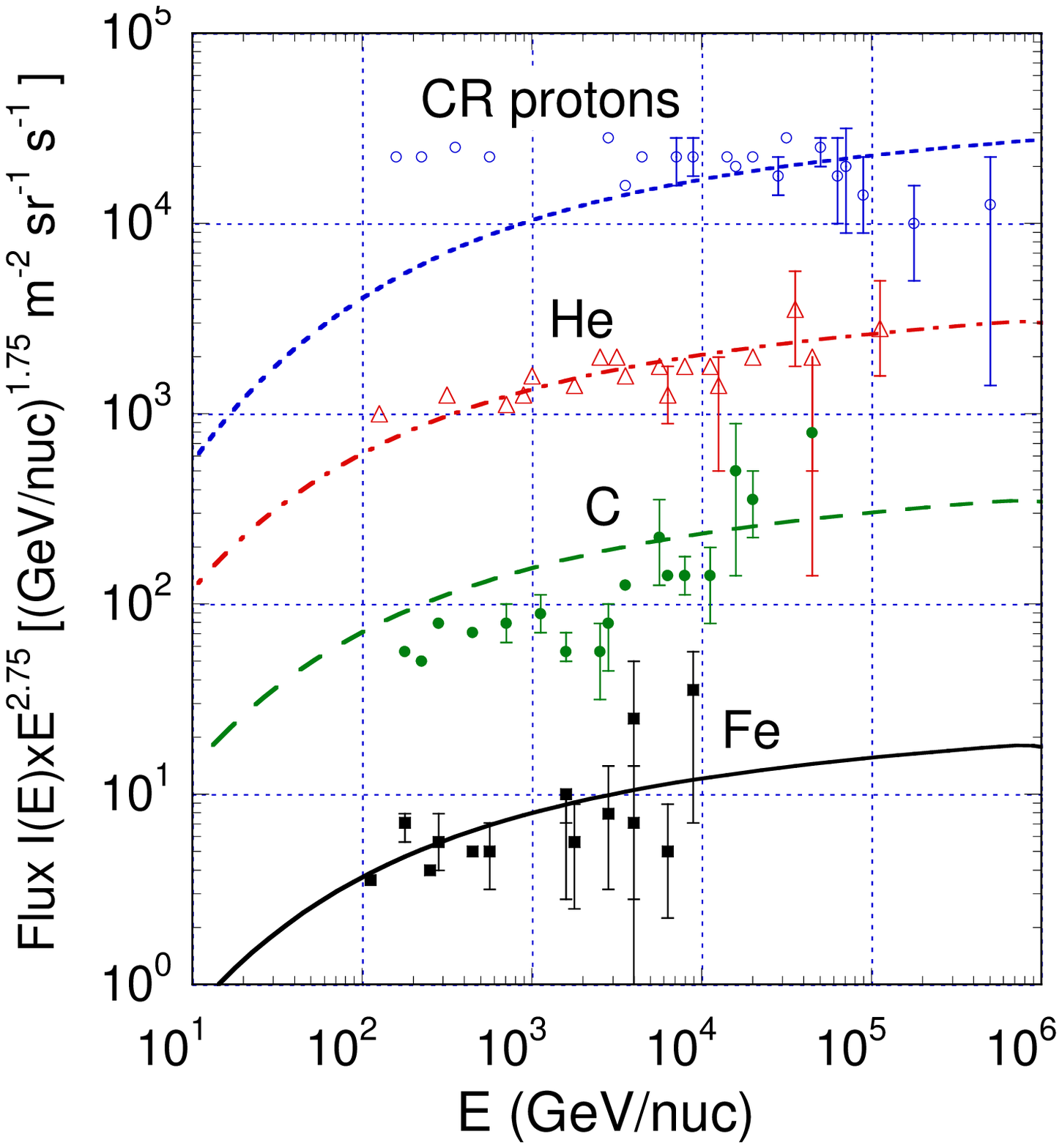}
}}
\caption{Contributions to the CR flux in the GeV/nuc - 
TeV/nuc range for a single
source with a low-energy cutoff determined by the minimum injection
momentum $(\beta\gamma)_{min} = 0.01$.  Data is taken from the papers
by Swordy et al.\ \cite{swo95,swo93}. The excess flux from the model,
especially for CR Carbon, suggest a low-enery cutoff in the CR
injection spectrum from GRBs.}
\label{fig:CR}
\end{figure}

This model must not overproduce CRs at energies below the knee. Fig.\
(\ref{fig:CR}) compares CR observations with a model that fits the
KASCADE data and has an injection momentum at $(\beta\gamma)_{min} =
0.01$. This model weakly overproduces the CR Carbon ions, though it is
not a serious discrepancy given the uncertainty in the KASCADE data
which determines the CR Carbon composition. A strong overproduction
would imply a low-energy cutoff in the energy of the injected cosmic
rays, or an earlier GRB.  A model with a low-energy cutoff at $10^5$
GeV would certainly be consistent with the medium-energy cosmic
rays. In the proposed scenario, SN contributions to CR production make
up the difference between CRs produced by a single GRB source and the
measured CR fluxes below $\approx 100$ TeV/nuc. The transition from
multiple SNs source to, primarily, a single source seems to occur at
energies of $\sim 1$ -- 100 TeV/nuc.

\subsection{Probability of Nearby GRB and CR Anisotropy}

We estimate the probability that we find ourselves in the midst of a
CR bubble formed by a recent, nearby GRB using the relation $ P \simeq
(r_{dif}/r_{MW})^2 N_{GRB} $. The quantity $(r_{dif}/r_{MW})^2$
measures the relative area of the Milky Way's disk covered by a CR
bubble with age $t$, and $N_{GRB}$ is the number of GRBs that would
have taken place during $t$. We normalize the rate of GRBs in the
Milky Way to $\dot N_{\rm M}$ GRBs per millennium, and $\dot N_{\rm
M}\approx 0.1$--0.3 (Section 2).  Thus $N_{GRB} = \dot N_{GRB} t =
10^3 \dot N_{\rm M}t_{\rm Myr}$.

Using model parameters, $r_{dif}= 2\sqrt{\lambda ct/3} \cong 4.9
t_{\rm Myr}^{1/2}(A\gamma_6/ZB_{\mu{\rm G}})^{1/6}$ kpc.  Most of the
active star formation in the Milky Way takes place within the spiral
arms found within $r_{MW} = 15r_{15}$ kpc. Thus $P \simeq 106
(t_{Myr}/r_{15})^2 \dot N_{\rm M} (A\gamma_6/ZB_{\mu{\rm G}})^{1/3}$,
and the probability is very unlikely when $P(\bar t_{Myr}) \ll 1$, or
when $t_{Myr}\lesssim \bar t_{Myr} \cong 0.1 r_{15} \dot N_{\rm
M}^{-1/2} $ $ (A\gamma_6/ZB_{\mu{\rm G}})^{-1/6}$.  The model fits use
a time of 0.21 Myr since the GRB exploded to fit the CR data, so there
is a reasonable probability for this event to have taken place, even
with $\dot N_{\rm M}\approx 0.1$.

CRs from a GRB event that occurred much earlier than 0.2 Myr cannot
fit the data unless the total energy release in CRs from a single GRB
is increased, which makes severe demands on GRB models (see
Fig.~\ref{fig:evolve}).  For the estimated values of the diffusion
coefficient, CRs from a GRB that occurred much later than $\approx 1$
Myr also makes severe energy demands on the GRB source. Thus a GRB
event occurring a few hundreds of millennia ago represents the most
probable situation.

The anisotropy $\omega =(I_{max}-I_{min})/(I_{max}+I_{min}) =
(3D/cn_{Z,A})(\partial n_{Z,A}/\partial r) =
(\lambda/n_{Z,A})(\partial n_{Z,A}/\partial r)$
\cite{gp76}. Substituting eq.~(\ref{nZA}) gives
\begin{equation}
\omega = {2 r \lambda\over r_{dif}^2} = 
{3\over 2}{r\over ct} \cong {0.2r_{500}\over t_{Myr}}\%.
\label{omega}
\end{equation}
The interesting point about eq.~(\ref{omega}) is that $\omega$ is
independent of energy for this diffusion model.

Analyses of the arrival directions of large numbers of CRs show
anisotropies near 3 PeV at the $\approx 0.15\pm 0.05$\% level
\cite{wat84,hil84}.  When $E$ is between 100 GeV and 100 TeV,
anisotropies of $\simeq 0.1$\% are found.  Schlickieser \cite{sch02}
discusses some sources of uncertainty in anisotropy measurements.  An
anisotropy of $\approx 1$\% is implied from eq.~(\ref{omega}) for
$t_{Myr} = 0.21$. If an anisotropy below $\approx 0.2$\% is confirmed,
then a number of implications follow. Either we are located near a
rather recent GRB, which could be unlikely, or the CR energy release
from GRBs is larger than given here. Moreover, there could be
contributions from a second GRB that would help isotropize the flux.
Indeed, CR contributions from a multitude of weaker SNe that do not
host GRBs could also help isotropize the CR flux. We discuss the
relative contributions of SNe to the CR spectrum in Section 7.

\section{Ultra-High Energy Cosmic Rays from GRBs}

\label{sec:x-gal}


In our model, the UHE component of the CR spectrum is produced by
extragalactic GRBs.  Throughout our calculations we take the local GRB
luminosity density in CRs to be $\dot\varepsilon_{CR} =
f_{CR}\dot\varepsilon_{GRB,X/\gamma} $ where
$\dot\varepsilon_{GRB,X/\gamma}$ is given in eqs.~(\ref{elldot}) and
(\ref{l44}), and $f_{CR}$ is the nonthermal baryon-loading fraction
required by a GRB model of HECRs.  Our fits to the data imply values
for $f_{CR}$.

\begin{figure}[c]
\centerline{\hbox{
\epsfxsize=400pt \epsfbox{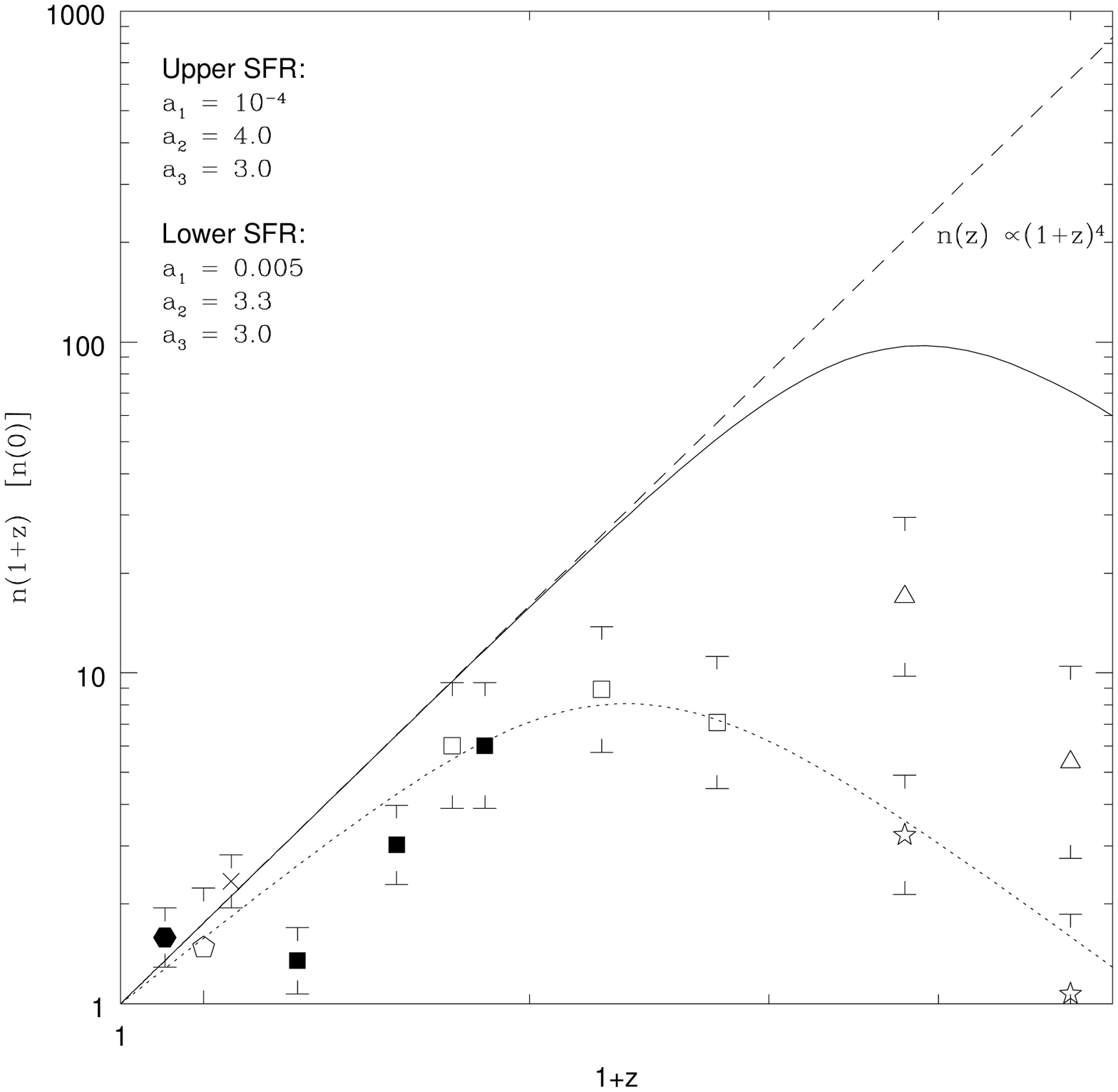}
}}
\caption{
The evolution of the rate density of GRBs as a function of $1+z$,
shown in comparison to measurements of the star formation rate (SFR)
inferred from measurements of restframe optical/UV emission from
galaxies.  The data, normalized with respect to a SFR of $0.09
~\rm{M}_{\odot}\rm{~Mpc}^{-3}\rm{~yr}^{-1},$ are listed in order of
increasing redshift and denoted for different measurements by the
following symbols: filled hexagons \cite{Gronwall}, open hexagons
\cite{Treyer}, diagonal crosses \cite{Tresse}, filled squares
\cite{Lilly}, open squares \cite{Connolly}, and open stars
\cite{Madau96}.  A SFR modification for dust extinction is given by
the open triangles \cite{Pettini}.  We use the function given in eq.\
(\ref{eq:evol}) with $a=0.005(0.0001)$, $b=3.3(4.0)$, and $c=3.0(3.0)$
for each of the models tested in our UHECR calculations.  The dotted
curve, based upon optical/UV observations, is arguably a lower limit
to the SFR evolution.  An extreme correction for dust extinction
\cite{Blain99} gives the stronger evolution shown by the middle, solid
curve.  For comparison, the dashed curve displays the relation
$n(z)=n(0)(1+z)^4$ used by other authors \cite{Scully,Berez}.  }
\label{fig:madau}
\end{figure}

We assume that the GRB cosmic rate-density evolution, in comoving
coordinates, follows the SFR history \cite{Madau} derived from the
blue and UV luminosity density of distant galaxies, with an analytic
fitting profile given by
\beq
n(z) = n(0)\;\frac{1+a_1}{(1+z)^{-a_2}+a_1(1+z)^{a_3}}~~
\label{eq:evol}
\eeq
\cite{der01}, where $n(z)$ is the GRB comoving rate density at epoch $z$. 
To accommodate present uncertainty in the true SFR evolution we take
two models with $a_1=0.005 (0.0001)$, $a_2=3.3 (4.0)$, and $a_3=3.0
(3.0)$ in line with the extreme ranges of optical/UV measurements
without (with) dust extinction corrections.  The lower optical/UV
curve is arguably a lower limit to the SFR evolution, while a much
stronger evolution of the SFR is found by Blain et al.\ \cite{Blain99}
after correcting for dust extinction.  To study the significance of
this uncertainty we consider a ``lower SFR'' and ``upper SFR'' in each
of our model fits.  We plot these SFR histories in
Fig.~\ref{fig:madau} in comparison with an oversimplified
$n(z)=n(0)(1+z)^4$ evolution used by other authors
\cite{Scully,Berez}, who perform similar UHECR calculations.   

Both our upper and lower SFR evolution differ significantly 
from the evolution $\propto (1+z)^4$, when $z\gtrsim 2$.   
For $>10^{20}$~eV CRs, this difference
is irrelevent as all of these CRs must have orginated from $z\ll 1$
due to severe CR attenuation during propagation, but the differing
SFRs have an important effect on the  predicted CR flux at energies 
$\lesssim 10^{18}$~eV.

Results from the Wilkinson Microwave Anisotropy Probe \cite{ben03}
and the High-$z$ Supernova Search Team \cite{Tonry} favor a 
$\Lambda$-cold dark matter ($\Lambda$CDM) cosmology
where $\Omega_{tot}=1.0,$~ $\Omega_{m}=0.3,$ and $\Omega_{\Lambda}=0.7$,
assuming a dark energy equation of state with $w=-1$ and
$h=0.7.$  We use this set of cosmological parameters 
throughout.  Our treatment of the UHECR flux
calculation, including attenuation during propagation, is presented
in Appenix A.

\subsection{KASCADE and HiRes Data Fits}

\label{sec:fit}

Our fits to the KASCADE and HiRes-I and HiRes-II Monocular data
cover the energy range from $\approx 2\times 10^{16}$~eV to
$3\times 10^{20}$~eV.  The low energy CRs
are from galactic GRBs (Section 3) and the high energy flux
is a superposition of extragalactic GRB sources.  
The region where the combined flux changes from halo-dominated
to extragalactic-dominated is near $E_{max}^{halo}.$
We carry out 8 different calculations where
in each case our minimum-$\chi^2$ routine fits the halo 
and extragalactic normalizations and finds the best-fit 
value for $E_{max}^{halo}.$   The value of $E_{max}^{halo}$ 
in turn fixes parameters in the halo propagation model.

We take characteristic values for the maximum cutoff energy from
acceleration in relativistic shocks of $E_{\rm{max}} = 10^{20}$~eV and
$10^{21}$~eV.  The spectral indices we consider are $p=2.0,$ where the
CR energy is injected equally over all decades in energy-space, and
$p=2.2$ where optimal fits are found in our Galactic diffusion model.
The upper and lower SFR evolution cases are studied in turn.  For each
case our model has three free parameters:
1) the CR halo-component cutoff energy $E^{halo}_{max}$; 
2)  the local CR luminosity density of extragalactic GRBs; and 
3) the relative intensity of the galactic halo CR component to the 
extragalactic component.
We also consider two values of the minimum energy $E_{min} =
10^{9}$~eV and $10^{14}$~eV for CR injection, which yield identical
fits but affect the overall CR energy requirements of a typical GRB.
For a soft $p=2.2$ spectrum, the baryon loading $f_{CR}$ is a factor
$\approx~(10^{14}/10^{9})^{0.2}=10$ lower for the $E_{min}=10^{14}$~eV
case than the $E_{min}=10^{9}$~eV case.
In total, we fit 43 data points covering the highest 4 decades of CR energies.
We estimate the quality of our fits with a reduced-minimum-$\chi^{2}$ routine
for 40 degrees of freedom.  


The transition region between the halo and extragalactic components
falls in the range $\sim 2 \times 10^{17}$~eV -- $2\times 10^{18}$~eV.
Thus, our data set must extend to low-enough energies to make a
determination of $E_{max}^{halo}$ practicable.  The lower energy data
we fit is from the KASCADE all-particle spectrum \cite{kam01},
starting at $10^{16.35}$~eV and continuing to $10^{17.15}$~eV.  The
HiRes-II Monocular data set covers $10^{17.25}$~eV to $10^{19.60}$~eV,
while the HiRes-I Monocular data covers the highest energies, from
$10^{18.55}$~eV to $10^{20.10}$~eV.
Our best-fitting model (marked with an asterisk in Table 2) was
considered in the context of the full KASCADE data set down to
energies of $\lesssim 10^{15}$~eV.

\begin{figure}[c]
{\hbox{
\epsfxsize=200pt 
\epsfbox{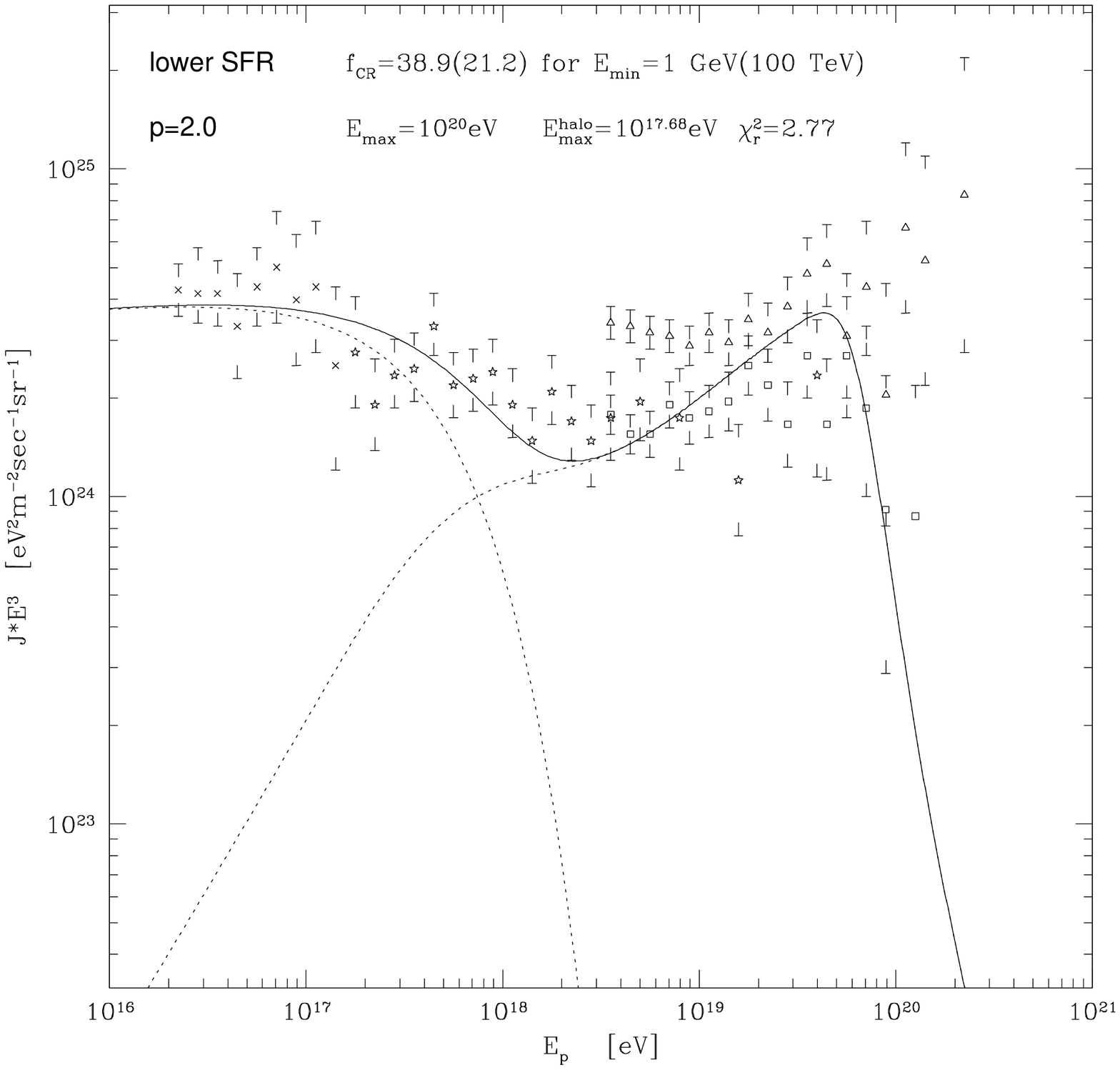}
\epsfxsize=200pt 
\epsfbox{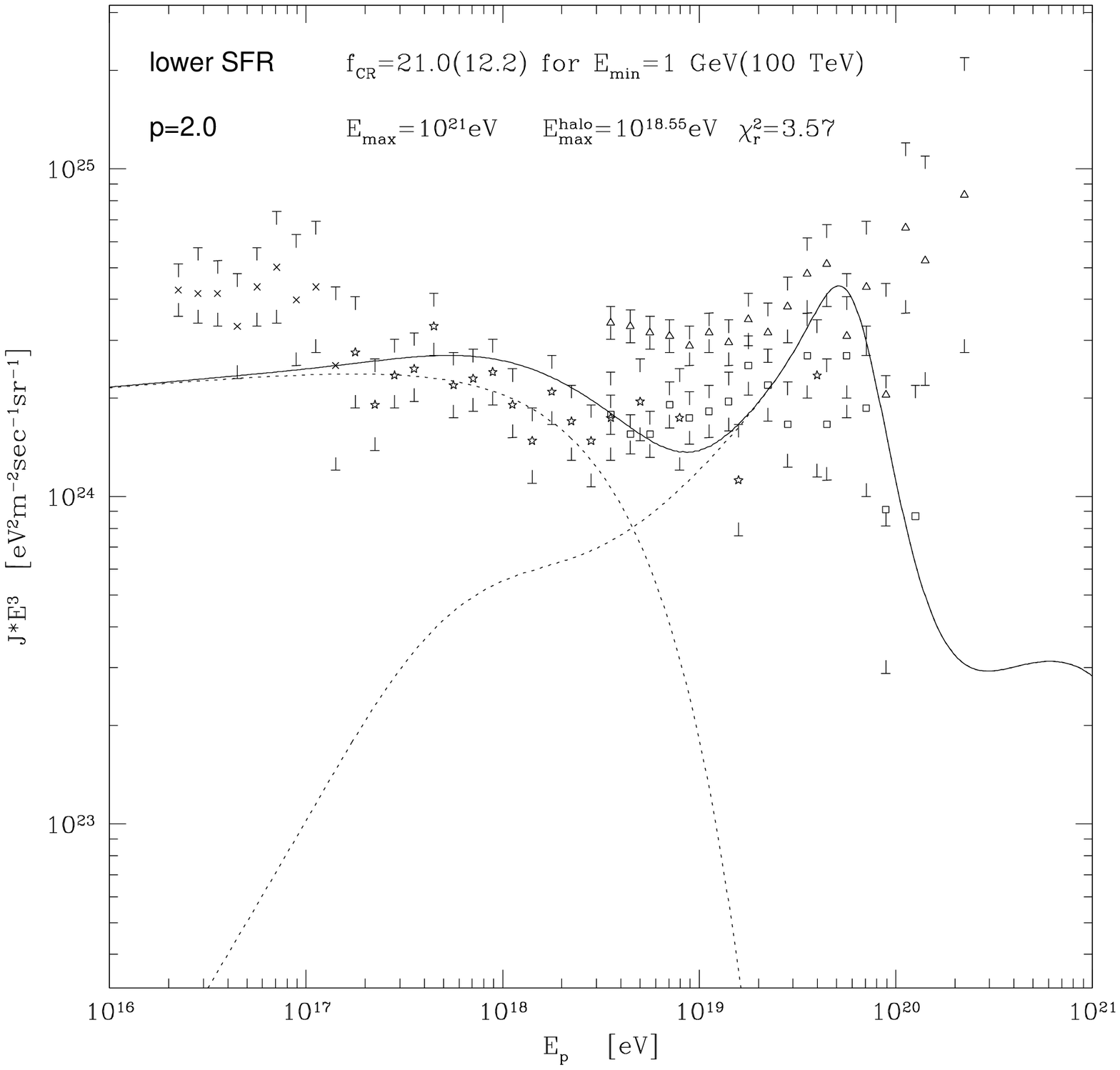}}
\vspace{0.5 in}
\hbox{
\epsfxsize=200pt 
\epsfbox{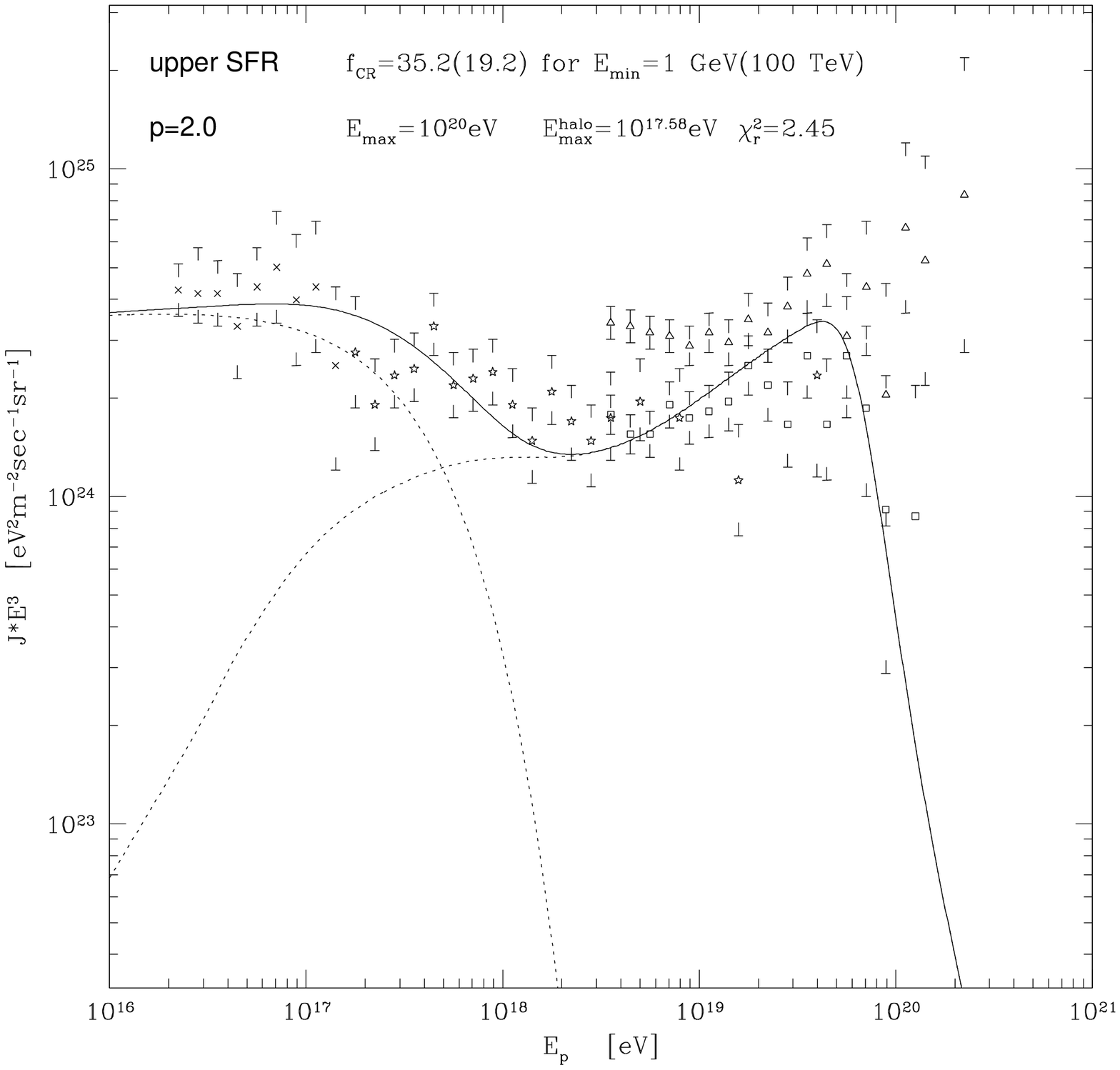}
\epsfxsize=200pt 
\epsfbox{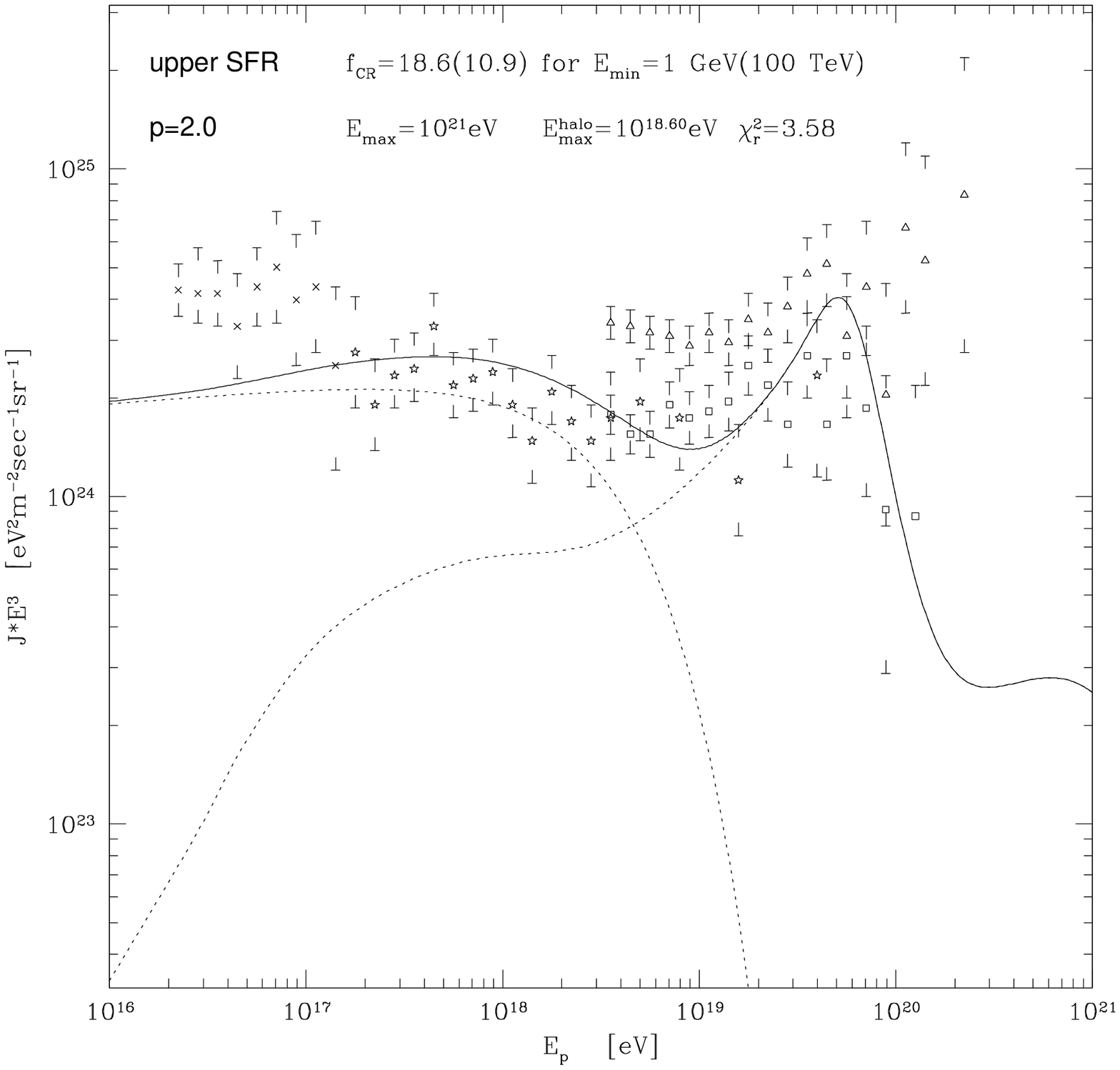}
}}
\caption{
Fits to UHECR data for model with injection spectral index $p=2.0$.
The fits to the KASCADE (crosses), HiRes-I Monocular (squares),
HiRes-II Monocular (stars) data assume various spectral cutoffs of the
source, and lower and upper SFR histories and $E_{max}$ as labeled on
the figures.  Although the AGASA data (triangles) are shown, they are
not included in the fits.  The cutoff energies for the halo component
$E_{max}^{halo}$, reduced $\chi^{2}_r$, and requisite local CR
luminosity densities $f_{CR}$ are given on each figure.  }
\label{fig:cr2}
\end{figure}

\begin{figure}[c]
{\hbox{
\epsfxsize=200pt 
\epsfbox{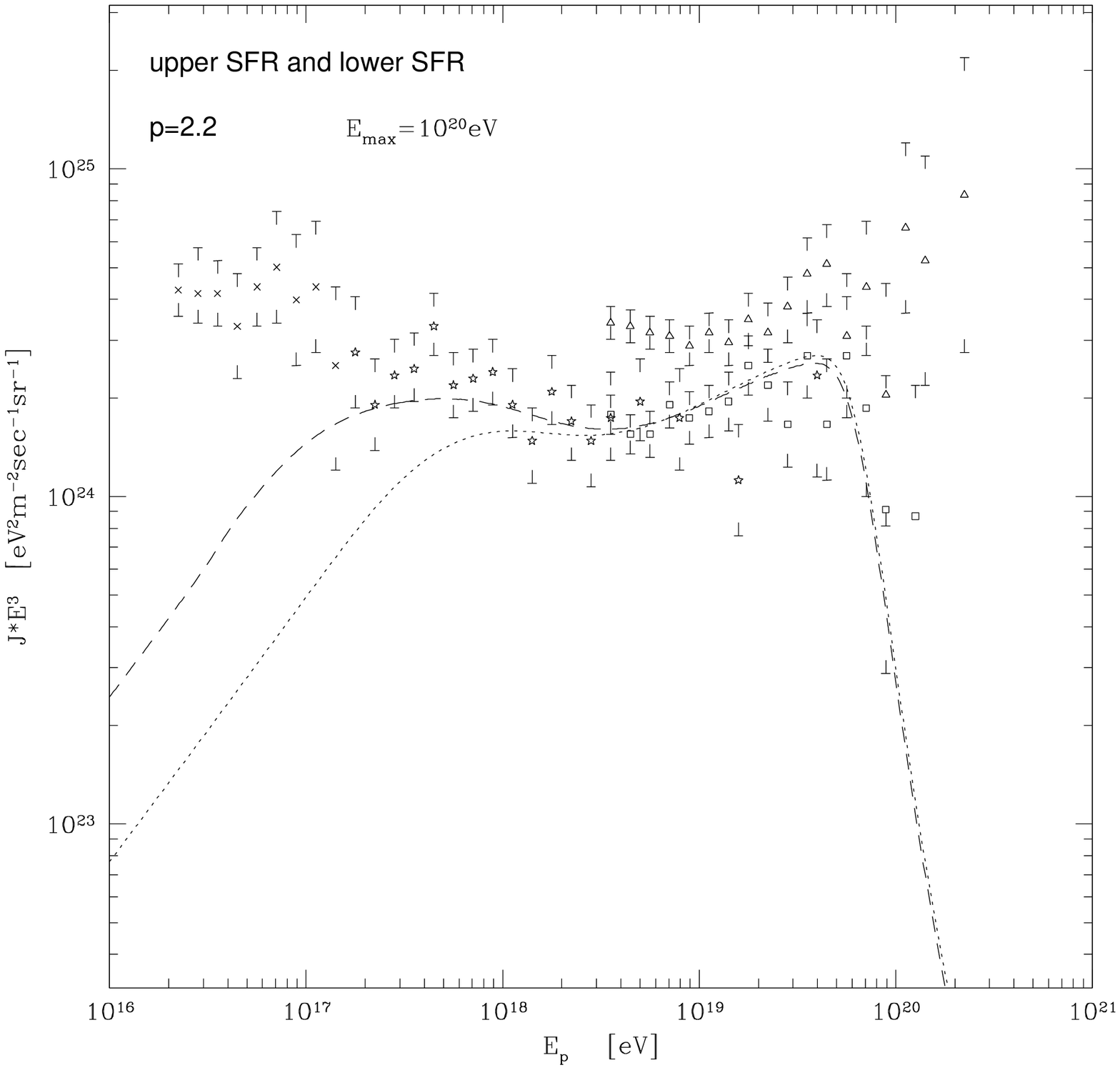}
\epsfxsize=200pt 
\epsfbox{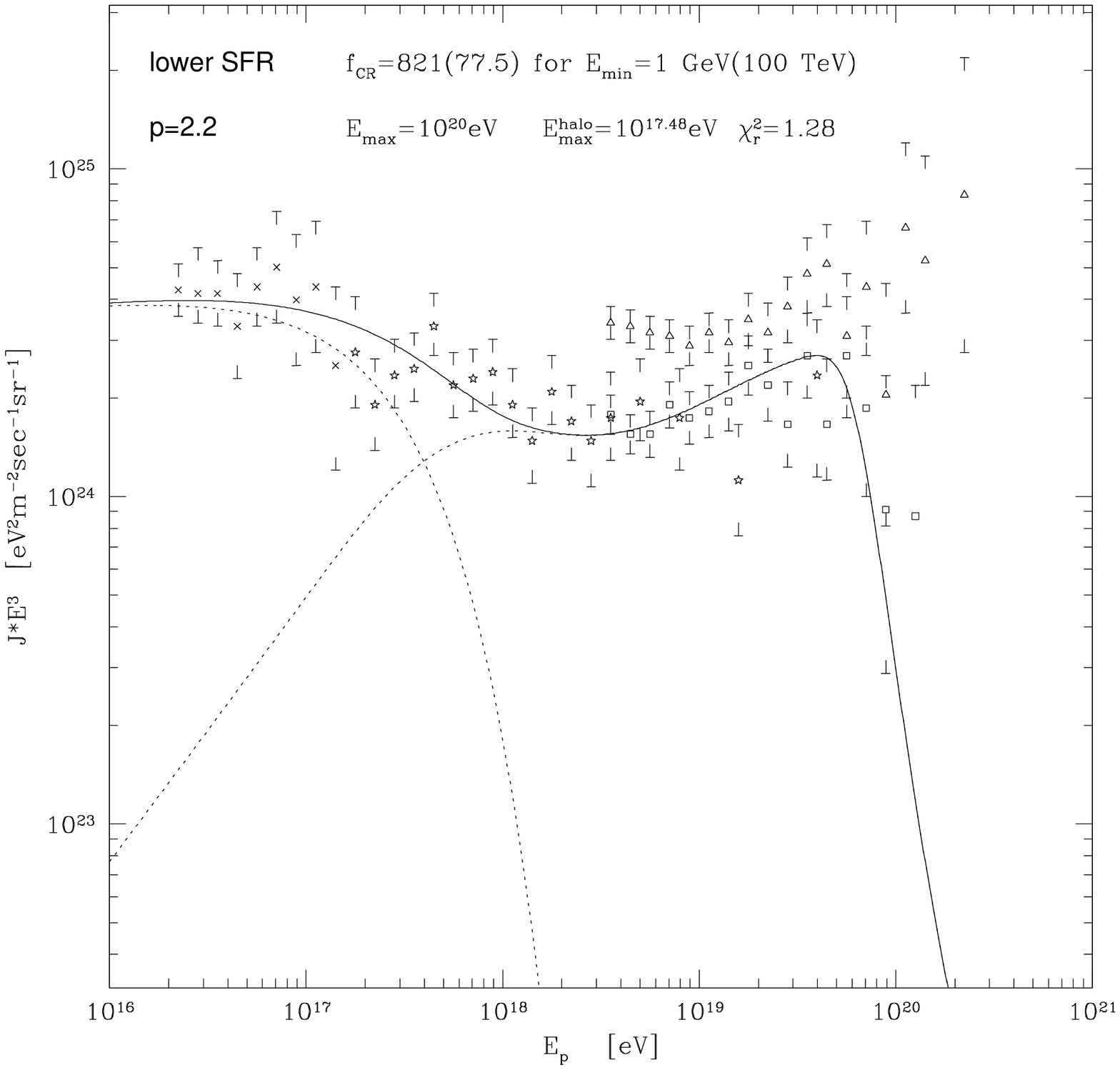}}
\vspace{0.5 in}
\hbox{
\epsfxsize=200pt 
\epsfbox{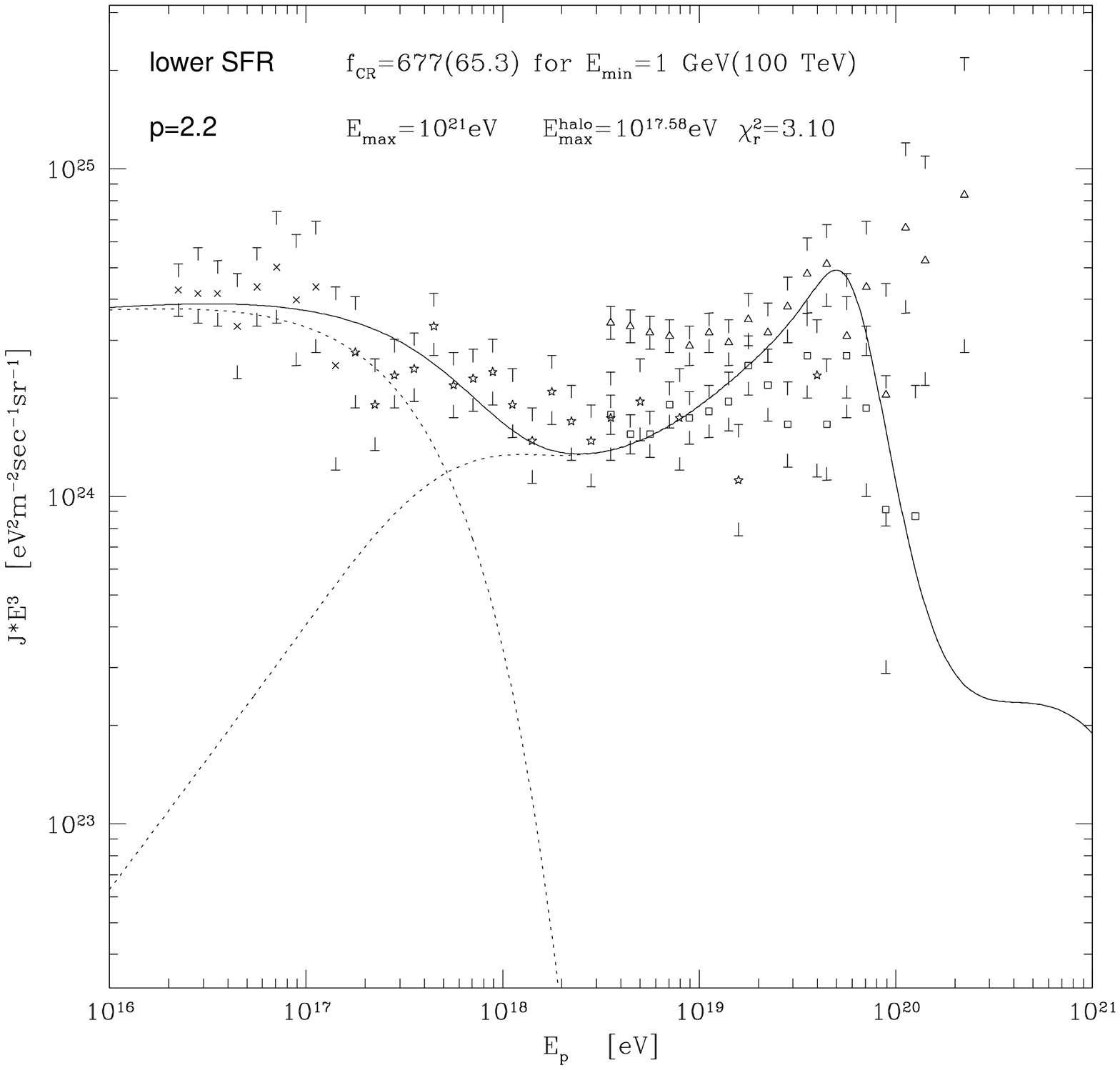}
\epsfxsize=200pt 
\epsfbox{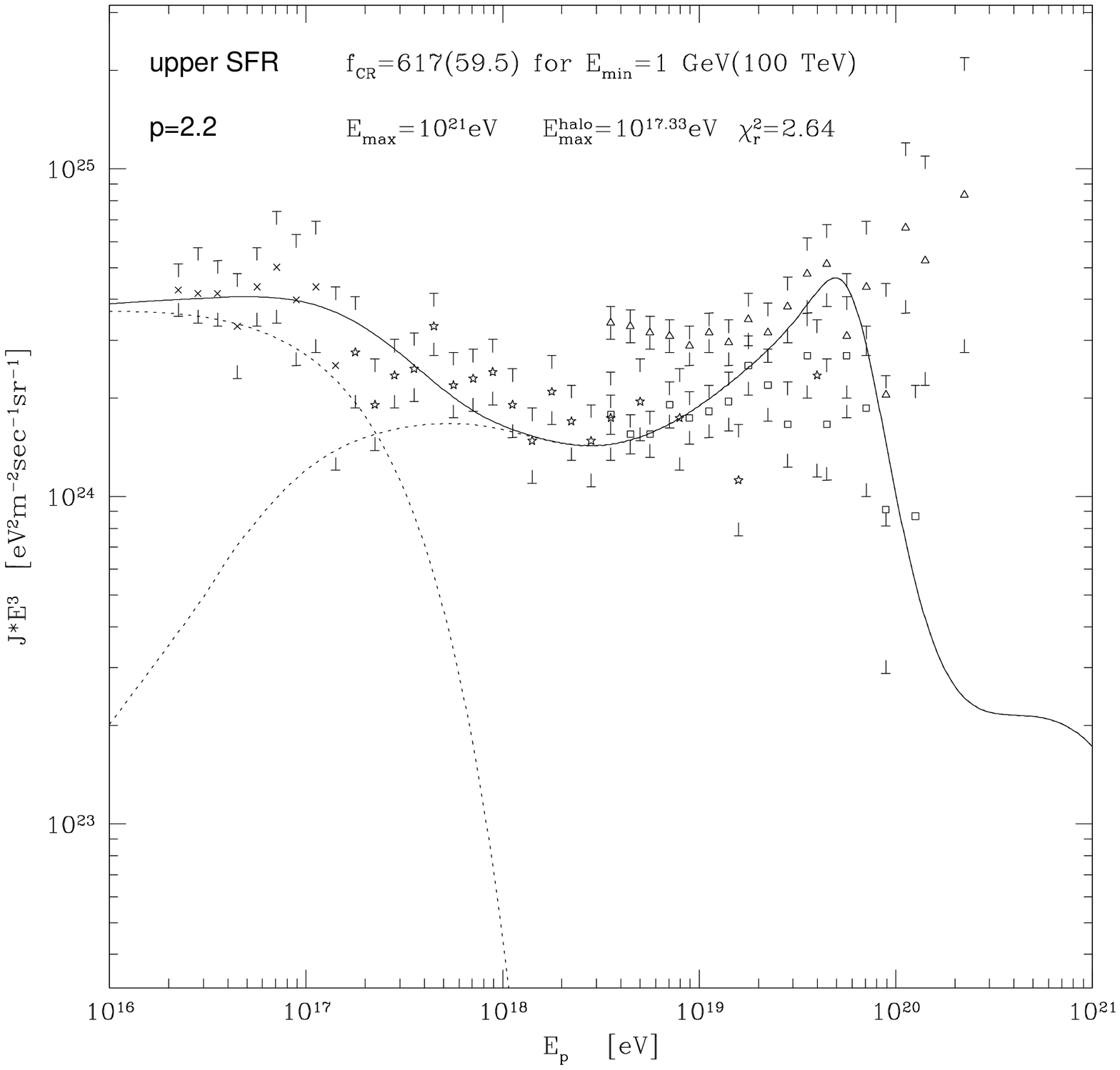}
}}
\caption{
Upper-left figure gives a comparison of the upper versus lower SFR
evolution.  Attenuation of the CR flux originating at energies $>
2\times 10^{18}$~eV and $z>1$ are swept by photo-pion and photo-pair
processes down to lower energies.  The increased number of CRs
injected for the upper SFR yields a significantly larger flux of CRs
in the range $2\times 10^{17}$~eV to $2\times 10^{18}$~eV.  The
remaining three figures have the same parameters as in Fig.\
\ref{fig:cr2}, except that the injection index is now $p=2.2$.  }
\label{fig:cr3}
\end{figure}

\begin{figure}[c]
\centerline{\hbox{
\epsfxsize=400pt \epsfbox{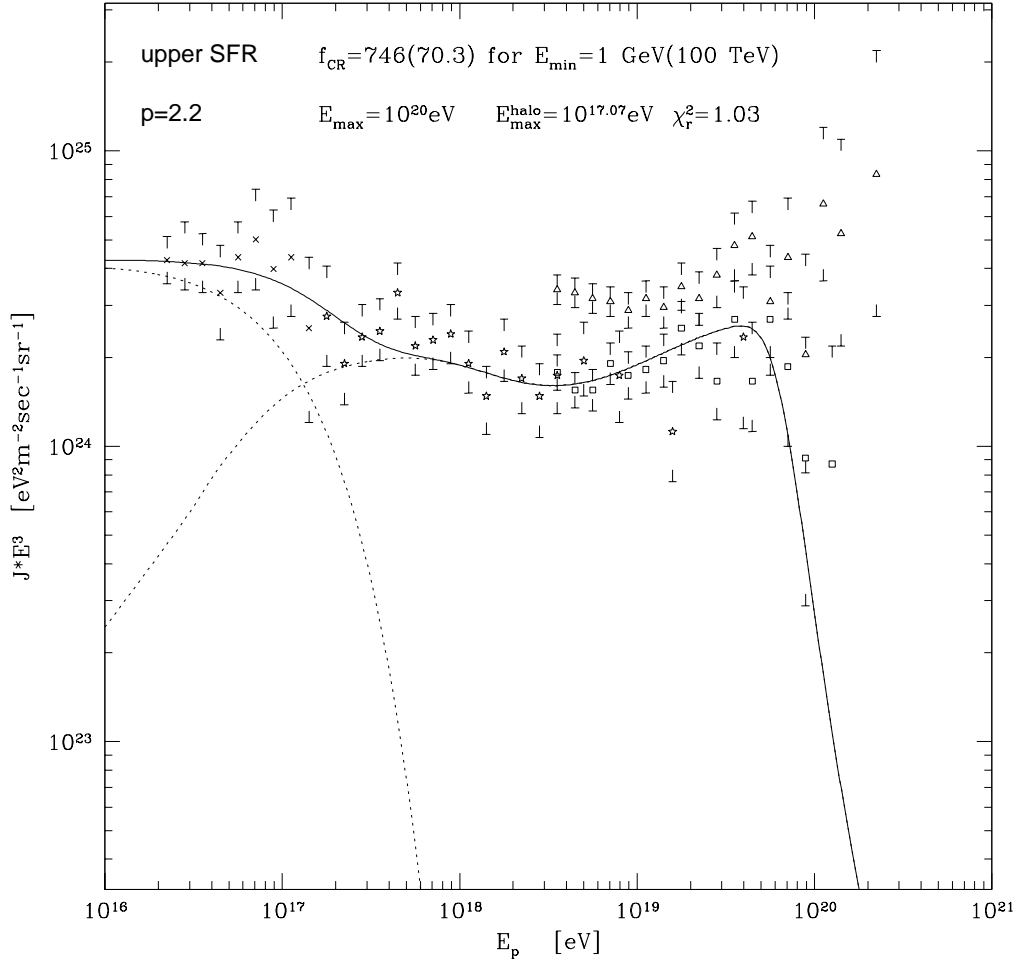}
}}
\caption{
Best fit to the Kascade (crosses), HiRes-I Monocular (squares),
HiRes-II Monocular (stars) data assuming a spectral cutoff at the
source of $E_{max}=10^{20}$~eV and using the upper limit to the SFR
evolution.  We also show the AGASA data (triangles) but do not include
these in our fits.  The cutoff energy for the halo component is
$E_{max}^{halo}=10^{17.07}$~eV giving a $\chi^{2}_r =1.03$ The
requisite baryon loading factor for this fit is $f_{CR}=746(70.3)$ for
a low energy cutoff at the source of $E_{min}=10^{9}(10^{14})$~eV.  }
\label{fig:cr4}
\end{figure}

The interplay between our model and features in the data is apparent
in the 8 calculations shown in Figs.~\ref{fig:cr2} -- \ref{fig:cr4}
and with the results listed in Table 2.  The steepening in the data
near the second knee ($10^{17.6}$~eV) is a feature which rules out
cases with $p=2.0.$ This assumes that the intercalibration between the
KASCADE and HiRes data is accurate, a conclusion that could be
modified if reanalysis of the HECR data reveal calibration
uncertainties. The extragalactic flux for $p=2.0$ spectra fall well
below the data at $E_{CR}< 10^{18}$~eV leaving the halo component
incapable of fitting the data both above and below the second knee.
Our best fitting set of model parameters is with $p=2.2$,
$E_{max}=10^{20}$~eV, and the upper SFR history.  This case puts the
transition between galactic and extragalactic CRs in the vicinity of
the second knee, consistent with evidence for a heavy-to-light
composition change in this energy range \cite{fly93}.  The ankle
energy (at $\approx 3\times 10^{18}$~eV) has a simple interpretation
as a suppression of the UHECR flux from the photo-pair process
(analogous to the GZK suppression from photo-pion production).

The extragalactic contribution to the CR flux in the $10^{17.6}$~eV --
$10^{18.6}$~eV range is driven by the strength of the SFR evolution.
The flux below $10^{18.6}$~eV is markedly higher in the upper SFR case
because the UHECRs produced at redshift $z$ are readily swept down to
$\sim 10^{18}/(1+z)$~eV by propagation effects (see Appendix A).  A
similar result was obtained \cite{Scully,Berez} in the case when the
cosmic evolution of GRB sources changes $\propto (1+z)^4$, which
approximates the upper SFR to $z\sim 2,$ but then diverges from the
upper bound to the SFR derived in Ref.\ \cite{Blain99}, hence
predicting too high a flux at $E_{CR}<10^{18}$~eV.

\begin{table*}
\caption{The results of various model fits to the KASADE, HiRes-I
and HiRes-II Monocular data. The parameters $p$ and $E_{max}$ give the
injection index and maximum injection energy of CRs from extragalactic
GRBs, respectively, for the upper and lower SFR curves given in Fig.\
\ref{fig:madau}.  The high-energy cutoff of the halo component
$E_{max}^{halo}$ and the required baryon load $f_{CR}$ are derived by
minimizing $\chi^{2}_{r},$ for low-energy CR injection cutoff
$E_{min}=10^{9} (10^{14})$~eV.  The best fit model is marked with an
asterisk. }
\vspace*{.3cm}
\label{GRBmodels} 
\begin{center}
\begin{tabular}{c c c c c c}
\hline 
Spectrum, $p$ & $E_{max}$(eV) & SFR evol. & 
$E_{max}^{halo}$(eV) & $\chi^{2}_{r}$ & $f_{CR}$  \\  
\hline 
2.2 & $10^{20}$ & lower & $10^{17.48}$  & 1.28 & 821(77.5) \\  
2.2 & $10^{21}$ & lower & $10^{17.58}$  & 3.10 & 677(65.3) \\  
2.2$^\ast$~ & $10^{20}$ & upper & $10^{17.07}$ & 1.03 & 746(70.3) \\  
2.2 & $10^{21}$ & upper & $10^{17.33}$  & 2.64 & 617(59.5) \\  
2.0 & $10^{20}$ & lower & $10^{17.68}$  & 2.77 & 38.9(21.2) \\  
2.0 & $10^{21}$ & lower & $10^{18.55}$  & 3.57 & 21.0(12.2) \\  
2.0 & $10^{20}$ & upper & $10^{17.58}$  & 2.45 & 35.2(19.2) \\  
2.0 & $10^{21}$ & upper & $10^{18.60}$  & 3.58 & 18.6(10.9) \\  
\hline 
\end{tabular}
\end{center}

\vspace*{.6cm}
\end{table*}

\subsection{AGASA Data Fits}

\begin{figure}[c]
\centerline{\hbox{
\epsfxsize=400pt \epsfbox{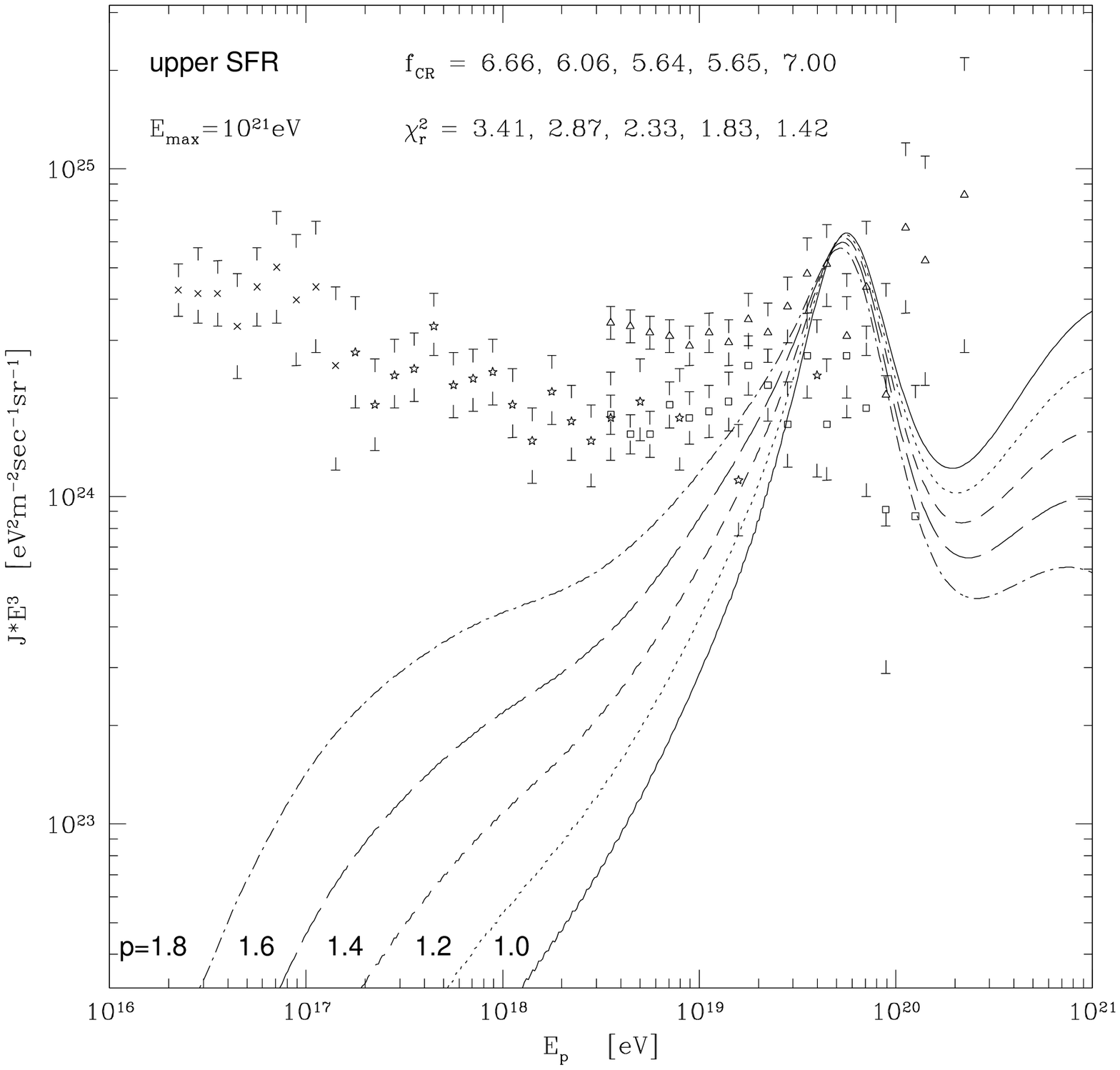}
}}
\caption{
Best fits to the highest nine energy bins in the AGASA data (tiangles)
$3\times 10^{19}$~eV -- $3\times 10^{20}$~eV for various hard spectra,
$p=1.0, 1.2, 1.4, 1.6, 1.8.$  The resultant $\chi^{2}_{r}$ and requisite
baryon loading $f_{CR}$ for each case
are shown on the figure. The KASACADE and HiRes data, although shown, are not
included in the fits.
}
\label{fig:crAGASA}
\end{figure}

The reported AGASA flux above $10^{20}$~eV provides a clue that UHECRs
may violate the GZK-cutoff.  We explore this possibility within the
context of an astrophysical source with a hard ($p<2.0$) spectrum and
cutoff $E_{max}=10^{21}$~eV sufficient to produce a super-GZK flux.
The results of five models are presented in Figure \ref{fig:crAGASA}
with fits to AGASA's highest nine energy bins.  For hard CR-spectra
GRB sources, the CR energetics are not a problem because the
high-energy portion of the spectrum receives most of the nonthermal CR
energy.  However, it is clear from our fits that if a super-GZK flux
is verified by observations with the Auger Observatory, it calls for
either a multi-component model of astrophysical sources (including
nearby sources) or new physics in the form of top-down (for recent
reviews, see Refs.\ \cite{topdown,topdown1}) or hybrid (for some
examples, see Refs.\ \cite{hybrid,hybrid1,hybrid2}) scenarios.

\section{High-Energy Neutrinos from Gamma-Ray Bursts}

The prediction of our model that the energy output in relativistic
protons and nuclei accelerated by relativistic shocks of GRBs can be
much larger than the total power inferred from the observed X-ray/MeV
radiation, $ f_{CR} \gg 1$, suggests a very important conclusion
concerning the question of detectability of GRBs by forthcoming
km-scale neutrino detectors like IceCube.

A recent study \cite{da03} of this question shows that, at best, only
from the brightest fraction of GRBs with neutrino fluence $\Phi_{\nu
,tot} $ in the prompt phase exceeding $10^{-4}$ erg cm$^{-2}$, would
it be possible to detect multi-TeV neutrinos by IceCube. This is a
general model-independent result which only assumes that the radiation
fluence derived from the X-ray/MeV $\gamma$-ray measurements in the
prompt phase of GRBs does not strongly underestimate the total
radiation output of GRBs (that could be missed at higher energies).
This result follows from the direct calculation of the number of muon
neutrinos $N_\nu $ to be expected for a detector like IceCube, with an
area $A_{d}= 10^{10} A_{10} \,\rm cm^2$, assuming a spectral fluence
of the neutrinos $\epsilon \Phi_{\nu}(\epsilon) = 10^{-4} \Phi_{-4}$
erg cm$^{-2}$.  For a spectrum of $\epsilon \Phi_{\nu}(\epsilon)$ with
the spectral index $\simeq 0$ (i.e., with index $p_\nu \simeq 2$ for
the differential number fluence $\phi_\nu (\epsilon)$), a simple
derivation give $N_\nu \simeq 1.2\Phi_{-4} A_{10} (1+{1\over 2} \ln
\epsilon_{14}^{-1})$ for $\epsilon_{14}<1$, and $N_\nu \simeq
1.2\Phi_{-4} A_{10} /\sqrt{\epsilon_{14}}$ for $\epsilon_{14} >1 $
(see \cite{da03}). Thus if the neutrino fluence is comparable to the
X/$\gamma$ photon fluence, detection of $\nu_\mu$ neutrinos could be
expected only from very powerful GRBs, at a fluence level
$\Phi_{-4}\gg 1$.

Note that $\Phi_{-4} = 1$ corresponds to the neutrino fluence $\approx
2.3 \times 10^{-4}$ erg cm$^{-2}$ integrated per each decade of
energy. Note also the decline in the number of neutrinos to be
expected per equal energy fluences per decade when $\epsilon_{14} \geq
1$. This is explained by the change in the detection efficiency
$P_{\nu\mu}$ of upward-going muon neutrinos at this energy (see
\cite{ghs95}), which can be approximated as $P_{\nu\mu} \cong
10^{-4}\epsilon_{14}^\chi$, where $\chi = 1$ for $\epsilon_{14}<1$,
and $\chi =0.5$ for $\epsilon_{14}>1$.  Therefore, since a typical
spectrum of neutrinos expected from GRBs due to photomeson
interactions of HECRs extends well beyond 100 TeV, detection of even a
single neutrino at multi-TeV energies would become likely only if the
total fluence in the $\nu_\mu$ would significantly exceed $10^{-4}$
erg cm$^{-2}$.

\begin{figure}[c]
\centerline{\hbox{
\epsfysize=150pt \epsfbox{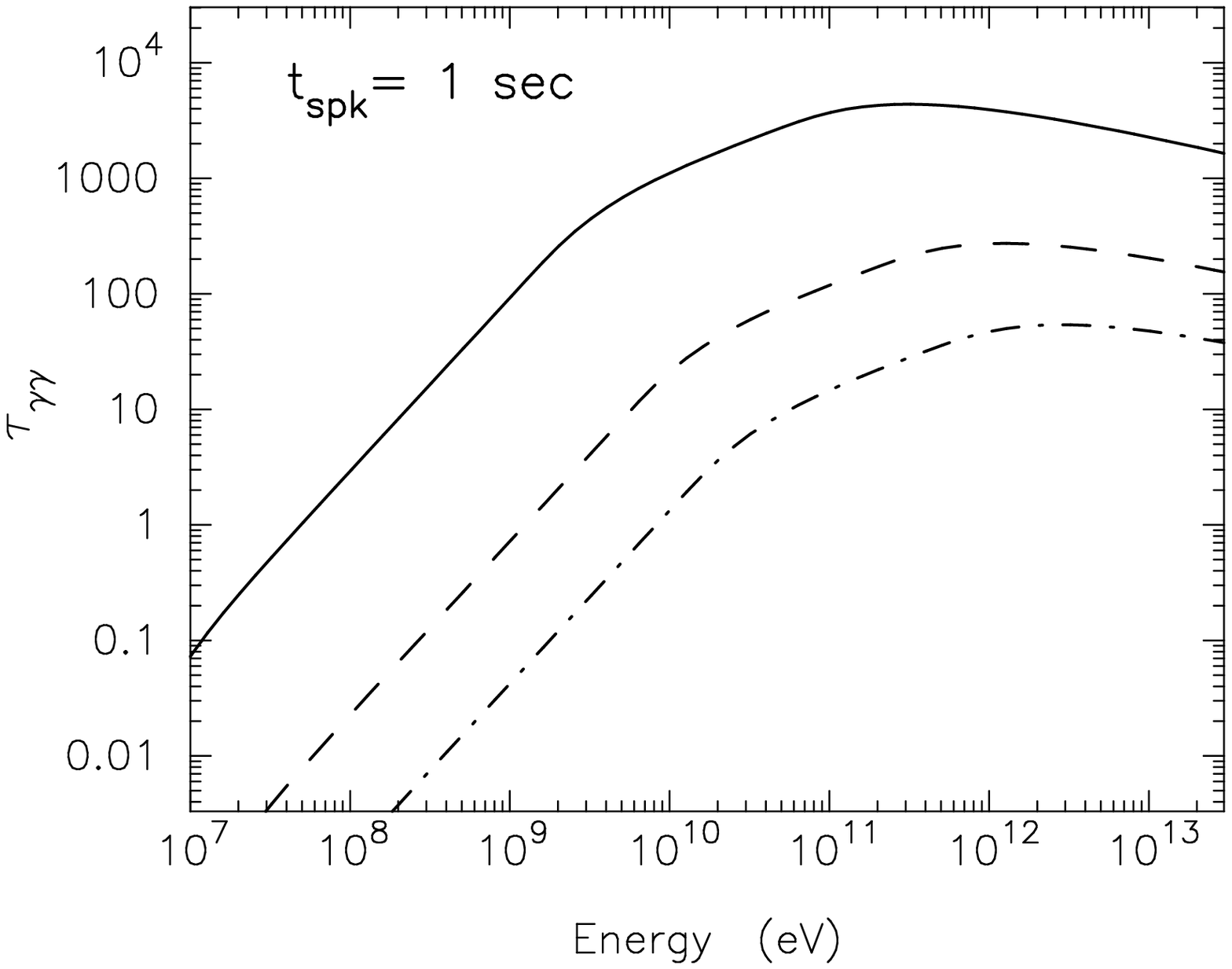} \hspace{2mm}
\epsfysize=150pt \epsfbox{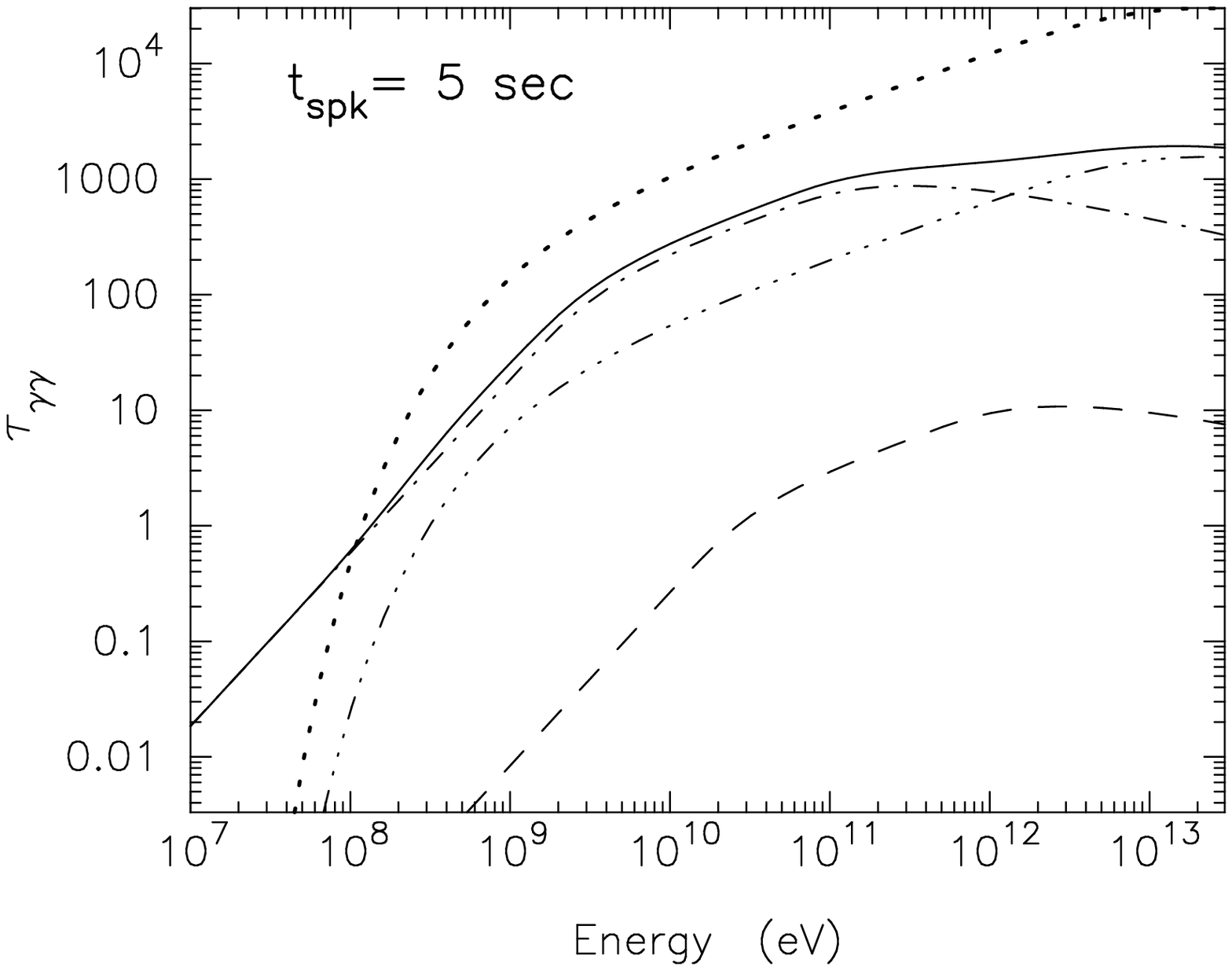}
}}
\caption{Opacity $\tau_{\gamma\gamma}$ for gamma-ray absorption in the 
radiation field inside individual spikes of a GRB at $z=1$, with the
X-ray/MeV gamma ray fluence $\Phi_{rad} = 3\times 10^{-4} \,\rm erg \,
cm^{-2}$ (with spectral flux peaked at 1 MeV) in the prompt phase with
a duration $t_{GRB}=100\,\rm s$, calculated for a collapsar GRB { \bf
(a)} ({\it left}): for 3 different Doppler factors $\delta = 100$
(solid curve), 200 (dashed curve), and 300 (dot-dashed curve),
assuming that the prompt emission is contributed by $N_{spk} = 50$
individual spikes with the mean duration $ t_{spk} = 1 \,\rm s$ each;
{ \bf (b)} ({\it right}): for $\delta = 100$ (dot-dashed curve) and
300 (dashed curve), but assuming $N_{spk} = 10$ spikes with the mean
duration $t_{spk}=5\,\rm s$. For comparison, here we also show
contributions to $\tau_{\gamma\gamma}$ due to the external radiation
field in the case of ``supranova" GRB with the assumed delay between
the supernova and GRB events $t_{sd} = 0.3 \,\rm yr$; the
3-dot--dashed curve corresponds to the opacity due to the external
photon field, and the solid curve shows the total
$\tau_{\gamma\gamma}$, across the spike in the case of $\delta =
100$. The dotted curve shows the opacity across the entire ``plerionic
nebula" with a radius $\simeq 4.6\, 10^{-3}\,\rm pc$ (see Ref.\
\cite{da03} for details of calculations in the ``supranova" scenario).
}
\label{tau}
\end{figure}

Meanwhile, the total amount of energy released in the neutrinos
(including $\nu_e$) is only about $40$-$50\,\%$ of the energy of
secondaries resulting from photomeson interactions. Therefore at least
the same amount of energy is released in the gamma-rays and electrons
of multi-TeV energies produced in the secondary pion decays.  For a
blastwave Lorentz factor $\Gamma \sim 100$-300, the GRB radiation
fields in the prompt phase are typically optically thick to
photo-absorption for gamma-rays with energies above the GeV domain as
is apparent from the curves in Fig.~\ref{tau}.  Even if one assumes
that the radiation in the prompt phase of a GRB is contributed only by
several individual spikes, with a characteristic duration as large as
$t_{spk} \sim 5$-10\,s, the source becomes transparent to gamma-rays
below the GeV domain only in case of a collapsar GRB with
Doppler-factor $\delta\gtrsim 300$.  Therefore most of the energy
injected in multi-TeV gamma-rays will be efficiently converted to hard
radiation in the X-ray to sub-GeV gamma-ray domains through the
synchro-Compton photon-pair cascade developing in the relativistic
shock/GRB source.

Our knowledge of high-energy ($\gtrsim 100$ MeV) emission from GRBs is
limited to 7 GRBs detected with the spark chamber on EGRET
\cite{bcs98} and the Milagrito detection of GRB 970417a
\cite{atk03}. The average spectrum of four GRBs detected
simultaneously with BATSE and the EGRET spark chamber from $\lesssim
100$ MeV to $\approx 10$ GeV is consistent with a $-2$ photon number
spectrum, implying that in these cases the $\gtrsim 100$ MeV fluence
does not exceed the BATSE fluence by more than a factor of a
few. However, the examples of the prompt and delayed high-energy
emission from GRB 940217 \cite{hur94}, the anomalous hard component
detected with BATSE and the EGRET TASC from GRB 941017 \cite{gon03},
and GRB 970417a, which required $\approx 10$ times more energy in the
TeV range than in the BATSE range, show that the high-energy behavior
of GRBs is yet poorly measured and even more poorly understood.  The
baseline assumption made here is to require that the energy fluence in
neutrinos not exceed the fluence observed in electromagnetic radiation
at hard X-ray and soft $\gamma$-ray energies.

Calculations of the neutrino fluxes expected from GRBs in case of
baryon-loading factor $f_{CR} \sim 1$, i.e., assuming equal energies
for relativistic hadrons and directly accelerated electrons (which are
assumed to produce the observed radiation), show that even in the case
of very powerful bursts with radiation fluence at the level
$\Phi_{rad} \sim 10^{-4}$ ergs cm$^{-2}$, which happen only
few/several times per year, the probability of detection of neutrinos
by a km-scale detector would remain hopelessly small in the framework
of the collapsar model unless the Doppler factor $\delta\lesssim 200$
or the variability time scale $t_{var} \ll 1$ s.  A significant
contribution to the target photon field density from a radiation
component external to the GRB source/blob, as in the plerionic
emission in the ``supranova'' scenario \cite{kg02}, or from radiation
scattered by the progenitor material in the circumburst environment in
the collapsar model, improves chances to detect neutrinos from GRBs
(see Ref.\ \cite{da03} for details). This is because for large Doppler
factors and $\Phi_{rad} \lesssim 10^{-4}$ ergs cm$^{-2}$, only a small
fraction of the injected proton energy can be converted into
secondaries through photomeson interactions of HECRs with internal
photons.

\begin{figure}[c]
\centerline{\hbox{
\epsfxsize=300pt \epsfbox{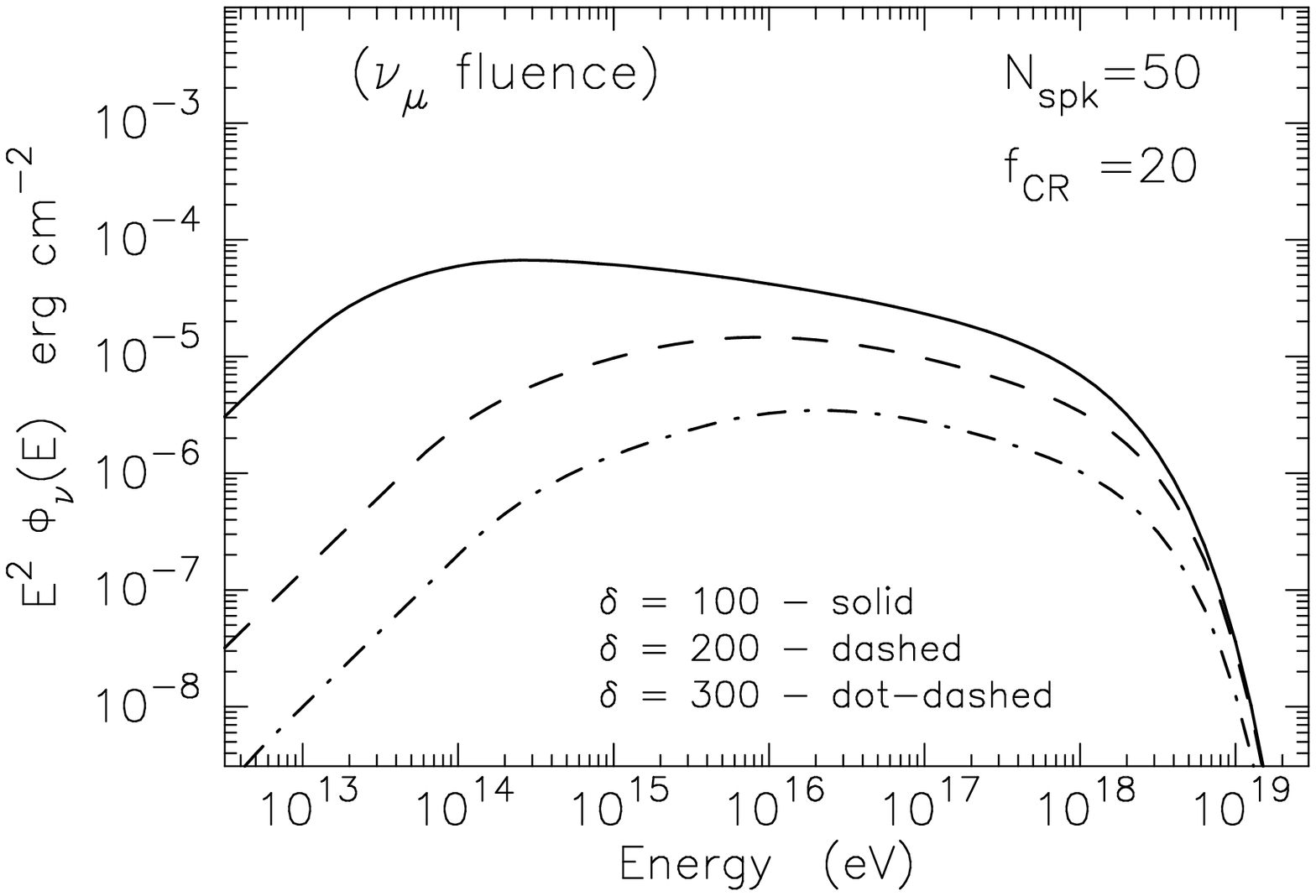}
}}

\vspace{2mm}
\centerline{\hbox{
\epsfxsize=300pt \epsfbox{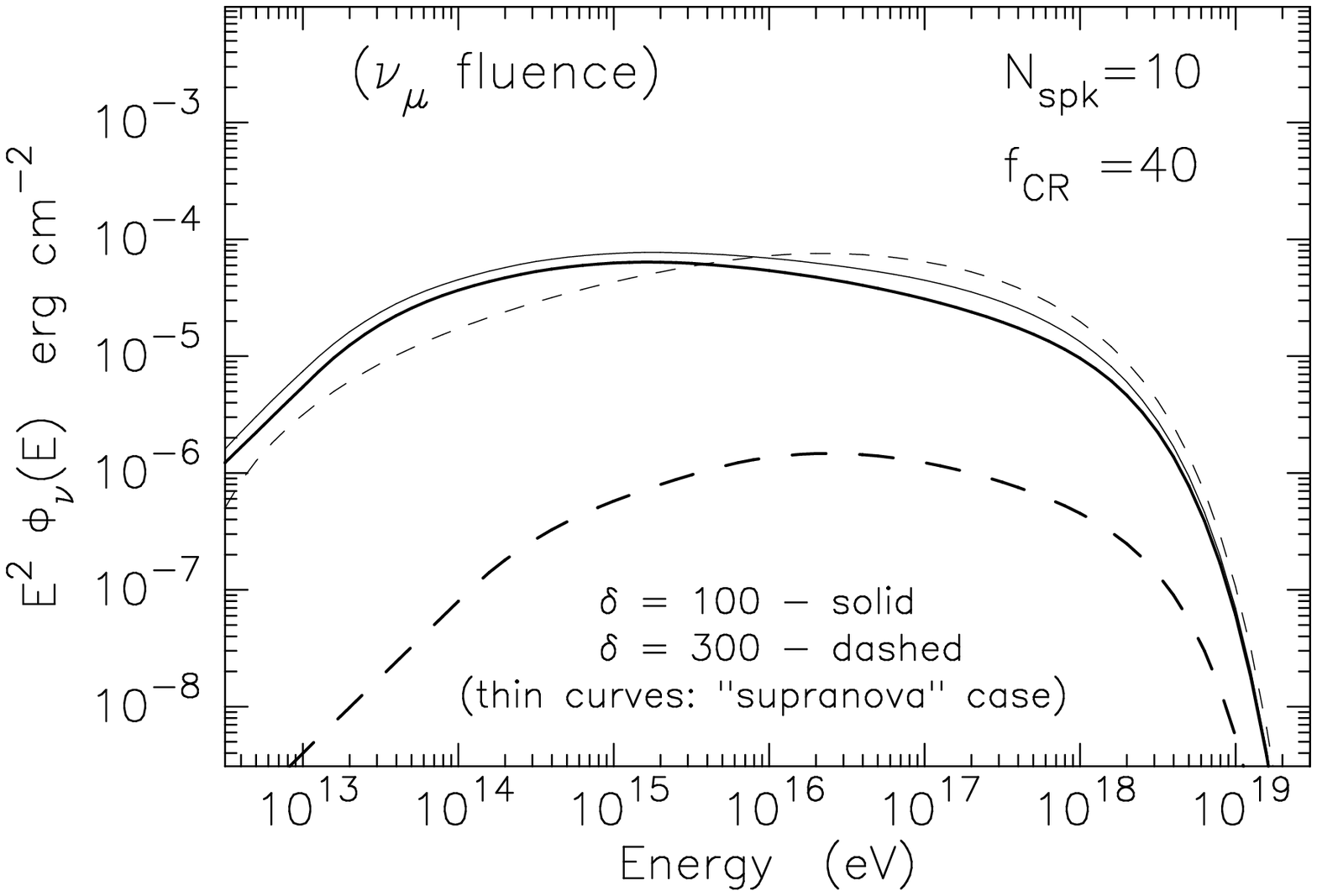}
}}
\caption{  The fluences of muon neutrinos calculated 
for a collapsar GRB assuming { \bf (a)} ({\it top}): 3 different
Doppler factors $\delta = 100$, 200 and 300, the same GRB parameters
as in Fig.~\ref{tau}a, and a nonthermal baryon-loading factor $f_{CR}=
20$; and { \bf (b)} ({\it bottom}): Doppler factors $\delta = 100$ and
$ 300$, but for the case of $t_{spk}=5\,\rm s$; for comparison, here
we also show by thin curves the fluences expected in the ``supranova''
model with the same parameters as in Fig.~\ref{tau}b.  } \label{neu}
\end{figure}

The current model for the origin of galactic high and ultra-high
energy CRs from GRBs requires that the nonthermal baryon loading
factor $f_{CR}\gg 1$, and this prediction will be tested by the
IceCube class neutrino detectors. In Fig.~\ref{neu}a we show the
neutrino fluxes expected in the collapsar scenario from a burst at the
radiation fluence level $\Phi_{rad} = 3\times 10^{-4}\,\rm erg\,
cm^{-2}$, calculated for 3 values of the Doppler factor $\delta$
assuming a GRB source at redshift $z = 1$ (for $h=0.65$). In order to
demonstrate clearly the dependence of the neutrino fluxes on $\delta$,
here we fixed the baryon-loading factor at a value $f_{CR} =20$ for 3
values of $\delta$. As in Fig.~\ref{tau}, for calculations in
Fig.~\ref{neu}a we assume that the prompt emission is contributed by
$N_{spk} = 50$ spikes with characteristic timescales $t_{spk} \simeq 1
\rm \, s$ each, which define the characteristic size (in the proper
frame) of individual spikes through $R_{spk}^\prime \simeq t_{spk}
\delta /(1+z)$.

The numbers of (muon) neutrinos to be expected by IceCube for $\delta
= 100,\, 200$ and 300 are $N_\nu = 1.32,\, 0.105 $ and 0.016,
respectively. We should note, however, that for the assumed value of
$f_{CR}$, the calculated total fluence of neutrinos (both $\nu_\mu$
and $\nu_e$) produced when $\delta = 100$ is $\Phi_{\nu ,tot} =7.2
\times 10^{-4}$ erg cm$^{-2}$, i.e., by a factor $7.2/3 = 2.4$ {\it
larger} than the assumed radiation fluence.  It means that in the
light of the arguments given above, the maximum value of the baryon
loading that could be allowed if the high-energy radiation fluence is
less than the X/$\gamma$ fluence for this particular case should be
about 8-10, instead of 20. Consequently, the expected number of
neutrinos for $\delta=100$ should be reduced to $\simeq 0.6$. On the
other hand, in Fig.~\ref{neu}a the neutrino fluence for the case
$\delta =200 (300)$ is equal to $\Phi_{\nu,tot} =1.4 \times 10^{-4} (3
\times 10^{-5}) \,\rm erg\, cm^{-2}$, so this accommodates a
baryon-loading increased from $20$ up to $f_{CR}\simeq 45 (200)$, with
the expected number of neutrinos observed by IceCube $N_{\nu ,corr}
\simeq 0.23 (0.16)$.  If the radiation fluence at MeV -- GeV energies
is allowed to exceed the X/$\gamma$ fluence by an order of magnitude,
a possibility that GLAST will resolve, then the expected number of
detected neutrinos could be increased correspondingly.

In order to understand the degree of stability of the predicted
numbers of neutrinos against the model assumptions, and most
importantly with respect to the assumed number of spikes which
determine the characteristic size/compactness of individual
``contributing blobs,'' in Fig.~\ref{neu}b we present the results of
calculations of the $\nu_\mu$ fluxes assuming $N_{spk}=10$,
corresponding to $t_{spk} \simeq 5\,\rm s$. In this case we chose a
higher $f_{CR} = 40$, anticipating a lower than previously considered
fraction of energy extraction from the relativistic protons. The heavy
solid and dashed curves show in Fig.~\ref{neu}b show the fluxes
calculated for $\delta = 100$ and 300, respectively, in the collapsar
scenario.  The corresponding total neutrino fluences are equal to
$6.9\times 10^{-4}$ and $1.3\times 10^{-5}\, \rm erg\, cm^{-2}$, and
the relevant numbers of $\nu_\mu$ to be expected for IceCube are
$N_\nu = 0.94 $ and $6.4\times 10^{-3}$. Correcting these numbers for
the maximum acceptable level of neutrino fluence assuming $\Phi_{\nu
,tot} \lesssim \Phi_{rad}$, results in $N_{\nu ,corr} \simeq 0.4$ and
$0.15$ for $\delta =100$ and 300, respectively, which are indeed close
to the numbers derived above in Fig.~\ref{neu}a. Note that for the
latter case, the implied baryon-loading factor would be as high as
$f_{CR} \simeq 900$, which could still be marginally acceptable, as
discussed in Section 5.1 above.

The two thin lines in Fig.~\ref{neu}b are calculated in the framework
of a ``supranova'' scenario for the same values of $\delta = 100$ and
300, assuming an external radiation field density in the ``plerionic
nebula'' with age $t_{sd}=0.3 \,\rm yr$ (see Ref.\ \cite{da03} for
details of calculations). The number of neutrinos to be expected when
$\delta = 100$ would be practically the same as for the collapsar
model, with $ N_{\nu ,corr} \simeq 0.5$. It is noteworthy, however,
that for $\delta = 300$, the predicted probability of neutrino
detection would be greatly increased relative to the collapsar model
prediction, namely $N_{\nu ,corr} \simeq 0.24$, which would at the
same time demand a significantly smaller value for the baryon-loading
factor, with $f_{CR} \simeq 15$ only.
  
Although these numbers are still smaller than 1, these calculations
leave a non-negligible probability for detection of 1 -- 2, and
hopefully up to a few, neutrinos from some of most powerful GRBs
during a reasonable observation time even in the case of a
``collapsar'' GRB with $\delta = 300$. Note that these neutrinos would
be mostly at energies $>100\,\rm{TeV}$ where almost no background
neutrinos are to be expected from a given direction to any single GRB
\cite{da03}.  Such a detection would be a confirmation of high
baryon-loading in GRB blast waves, and would also suggest a very
significant contribution of the accelerated hadrons in the observed
hard radiation through secondaries produced in the photomeson
interactions.

Another prediction of a hadronically dominated GRB model is the
possible detection of gamma-rays in the multi-GeV energy domain by
GLAST at a stage near the prompt phase when the density of radiation
fields would still be sufficient for extraction of a small fraction of
the energy of accelerated protons, but when the secondary gamma-rays
would not be suppressed by the photo-absorption process.  In case of
relatively close GRBs, with $z\ll 1$, these gamma-rays could then be
detectable also by ground-based gamma-ray detectors like VERITAS and
HESS with energy thresholds $\sim 100$ GeV.

\section{Discussion}

We have proposed a model where HECRs originate from GRBs.  The CR flux
near the knee is assumed to result from CRs produced by a single GRB
which has occurred relatively recently and not very far from us in the
Milky Way.  These CRs propagate diffusively in the Galactic disk and
halo.  The simple diffusive propagation model developed in Section 3
implies that the measured CR flux results from the modification of the
injection spectrum of an impulsive CR source due to transport through
a magnetic field with a given turbulence spectrum.  Using a turbulence
spectrum harder at smaller wavenumbers and steeper at larger
wavenumebrs, we have fit the 2001 KASCADE data for CR ion spectra
between $\approx 1$ and 100 PeV, and explained the change in the
all-particle spectra from $p=2.7$ to $p\cong 3.0$.  A GRB releasing
$\approx 10^{52}$ ergs in HECRs, located $\approx 500$~pc away, and
occurring $\approx 2\times 10^{5}$ years ago, provides reasonable fits
to the KASCADE data.

Our model of a single GRB source making CRs at energies through the
knee of the CR spectrum bears some similarity to the single-source
model proposed by Erlykin and Wolfendale \cite{ew02} to fit data near
the knee. These authors argue, however, that propagation (``Galactic
modulation") effects cannot explain the constant rigidity break of the
CR ionic species, whereas we employ a propagation model to produce
that break. In this respect our propagation model treats
rigidity-dependent transport as in the model of Swordy \cite{swo95}
and builds upon the detailed study of Atoyan, Aharonian, and V\"olk
\cite{aav95} for the spectral modification effects due to
energy-dependent diffusive propagation of CRs from a single source.
Our model explains, moreover, CR data not only through the first and
second knees but also at the highest energies.

The turbulence spectrum that fits the CR spectrum near the knee
employs a Kraichnan spectrum at small wavenumbers and a Kolmogorov
spectrum at large wavenumbers.  Turbulence is thought to be generated
at the smallest wavenumbers or largest size scales, with subsequent
energy cascading to smaller size scales \cite{clv02}.  It is
interesting to note that there are two crucial length scales in our
turbulence spectrum, namely $k_0^{-1} \approx 100$ pc, and $k_1^{-1}
\approx 1$ pc. The generation of turbulence in the disk and halo of
the galaxy at the larger size scale could be associated with halo-disk
interactions (e.g., through the interactions of high-velocity clouds
with the Galactic disk), which would deposit turbulence throughout the
disk and the halo of the Galaxy on a size scale $h_d \approx 100$
pc. The smaller length scale is typical of the Sedov length scale for
supernova explosions in the disk of the Galaxy. Indeed, SNe would
generate a large amount of turbulence energy which could make a
distinct contribution to the turbulence spectrum in this wavenumber
range.

The origin of the different indices of the two components of the MHD
turbulence spectra at small and large wavenumbers could be related to
the time available for the turbulence energy injected at the different
size scales to cascade to larger wavenumbers.  Medium-energy CRs with
energies between $10^{9}$ -- $10^{14}$ eV/nuc will diffuse by
gyroresonant pitch-angle scattering in response to MHD waves with
$k\gg 10$/pc. The model turbulence spectrum at large values of $k$ is
given by a Kolmogorov spectrum with index $q = 5/3$, as seen in Fig.\
(\ref{fig:kwk}).  Because medium-energy CRs are thought to arise from
a superposition of many SNe, we can treat their transport in the
framework of continuous injection.  As noted previously, the measured
index of CRs from continuous sources is steepened by a factor $2-q =
1/3$. If medium-energy CRs, whose measured number intensity index is
$\approx 2.7$ \cite{sim83}, result from many SNe that produce CRs
which diffuse through pitch-angle scattering in a spectrum of MHD
turbulence with index $q = 5/3$, then it is necessary that the
injection indices of these medium-energy CRs lie between $\approx 2.3$
and 2.4.

This is a surprising result, because it is generally thought that the
strong shocks in SNe accelerate and inject CRs with an injection
closer to 2.0 than 2.4 \cite{kir94}.  Such a soft injection spectrum
could be avoided if medium-energy CRs diffuse in a turbulence spectrum
with index $q = 3/2$. In this case, the measured index is steepened by
0.5 units compared to the injection index. If the large-$k$ component
with the steeper index is superposed on the extrapolation of the
small-$k$ component to large wavenmbers, then at sufficiently large
values of $k$, the turbulence spectrum will change from a Kolmogorov
to a Kraichnan spectrum, as illustrated by the long-dashed line in
Fig.\ \ref{fig:kwk}.  In this case, a continuous injection scenario
implies that an injection index of 2.2 yields an observed CR spectrum
with a 2.7 number index. A $p = 2.2$ injection index is in accord with
expectations from nonrelativistic shock acceleration. Note that
wavenumbers of 10$^{3}$ pc$^{-1}$ are gyroresonant with $\sim 30$ TeV
CR protons.

An important feature of recent GRB studies is their association with
SNe.  CR acceleration at SN or GRB shocks is crucial for the
production of CRs from the lowest energies, $\lesssim$ GeV/nuc, to the
highest energies $\gtrsim 10^{20}$ eV. The different speeds of the SN
shocks in the different types of SNe ejecta ranging from relatively
slow Type II ($\approx 3000$ -- 10000 km s$^{-1}$) to marginally
relativistic Type Ib/c is important to produce the full cosmic ray
spectrum \cite{der01a,sve03}. We therefore predict that HESS and
VERITAS, with their improved sensitivity and imaging, will detect
$\gamma$-ray emission from supernova remnants at a low level unless
that SN has also hosted a GRB.

At energies $E \gtrsim E_{max}^{halo}\approx few \times 10^{17}$~eV,
CRs stream out of galaxies to form the metagalactic CR component.  We
assume that GRBs evolve with cosmic epoch according to the SFR history
of the universe, so that most of the UHECRs are produced at redshift
$z\gsim 1$ when the SFR is greatest.  The UHECRs have their spectrum
attenuated by photo-pion, photo-pair, as well as adiabatic energy
losses which become important for particle energies below $2\times
10^{18}/(1+z)$~eV.  Our best fit was found to have spectral index
$p=2.2$ and $E_{max}=10^{20}$~eV.  For these parameters, a slightly
better fit was found for the stronger GRB redshift evolution (``upper
SFR'').  Stronger GRB evolution contributes more of the extragalactic
CR flux over the range $2\times 10^{17}$~eV to $2\times 10^{18}$~eV
than the lower SFR case, giving a best-fit value $E_{max}^{halo} \sim
2\times 10^{17}$~eV.  This is an effect of CR attenuation from
photopion and photopair production processes where all of the CR flux
produced at $E\gsim 2\times 10^{18}$~eV (and $z\gsim 1$) are swept to
lower energies.  Crucial for deriving these conclusions are the
relative calibrations of the KASCADE and High-Res experiments.

Our best fit to the UHECR data gives a measure of the local GRB
luminosity density $\dot\varepsilon_{CR}$ required in CRs.  We find
$f_{CR}\approx 70$ for $E_{min}=10^{14}$~eV and $\delta=300$.  This
implies that GRBs must be baryon-loaded by a factor $\gtrsim 50$ if
this model for HECRs is correct. The precise value of $f_{CR}$ varies
with Doppler factor according to $f_{CR}\approx 70(300/\delta)^{0.4}$,
for a $p = 2.2$ injection spectrum.  Nonthermal baryon-loading factors
$f_{CR} \gg 1$ implies that GRBs should also be luminous in
high-energy neutrinos.  In section 6 we calculate the fluence in
neutrinos and predict that $\approx 0.1$ -- few neutrinos of energy
$100$~TeV--$100$~PeV\ may be observable in IceCube a few times per
year from individual GRB explosions, which depends on the radiation
fluence measured from GRBs at $\gtrsim 100$ MeV -- GeV energies, which
is not yet well known.

In a collapsar model calculation \cite{da03}, the predicted number of
neutrinos depends sensitively on $\delta$, because the density of
internal synchrotron photons is much larger for smaller values of
$\delta$. The X-ray flashes \cite{Heise} may be ``dirty fireball"
GRBs, which are similar to classical long-duration GRBs, though with
larger (thermal) baryon loading and smaller average values of $\delta$
\cite{dcb99}.  This suggests that the X-ray flashes may be more
promising candidates for neutrino detection than GRBs with peak photon
energies of the $\nu F_\nu$ flux at several hundred keV energies.

The low statistics of current data at the highest energies make
unclear the presence or absence of a super--GZK UHECR component.  The
HiRes data appears to be consistent with the GZK effect, thereby
favoring an origin of UHECRs in ``conventional" astrophysical sources.
If the UHECR flux is found to exceed predictions of a GZK cutoff, then
this could be evidence of new physics, could indicate the existence of
super-GZK sources of cosmic rays within $\sim 50$ Mpc from Earth, or
both.  Our current interpretation of the data is made uncertain by
results from the AGASA experiment \cite{tak98} which reveals a
super-GZK flux, in conflict with the HiRes and Fly's Eye data
\cite{hires} (see, however, Ref.\ \cite{mbo03}, who argue that the
discrepancy is not serious given the uncertainties in calibration and
the statistical variance of CR fluxes measured at such high energies).
 
The possibility of new physics associated with a super-GZK CR flux has
generated much interest.  Scenarios for a super-GZK flux include the
Z-burst scenario \cite{hybrid}, magnetic monopole primaries
\cite{hybrid1,hybrid2}, and the decay of supermassive relic
particles\cite{topdown}.  (For a recent review, see Ref.\
\cite{topdown1}.)  Another approach to generate super--GZK fluxes is
from astrophysical sources distributed within a GZK--distance from
Earth, $\lsim 140$~Mpc. The Auger detector should resolve this
question by measuring $\approx 30$ events per year above $10^{20}$~eV.

\begin{figure}[c]
\centerline{\hbox{
\epsfxsize=400pt \epsfbox{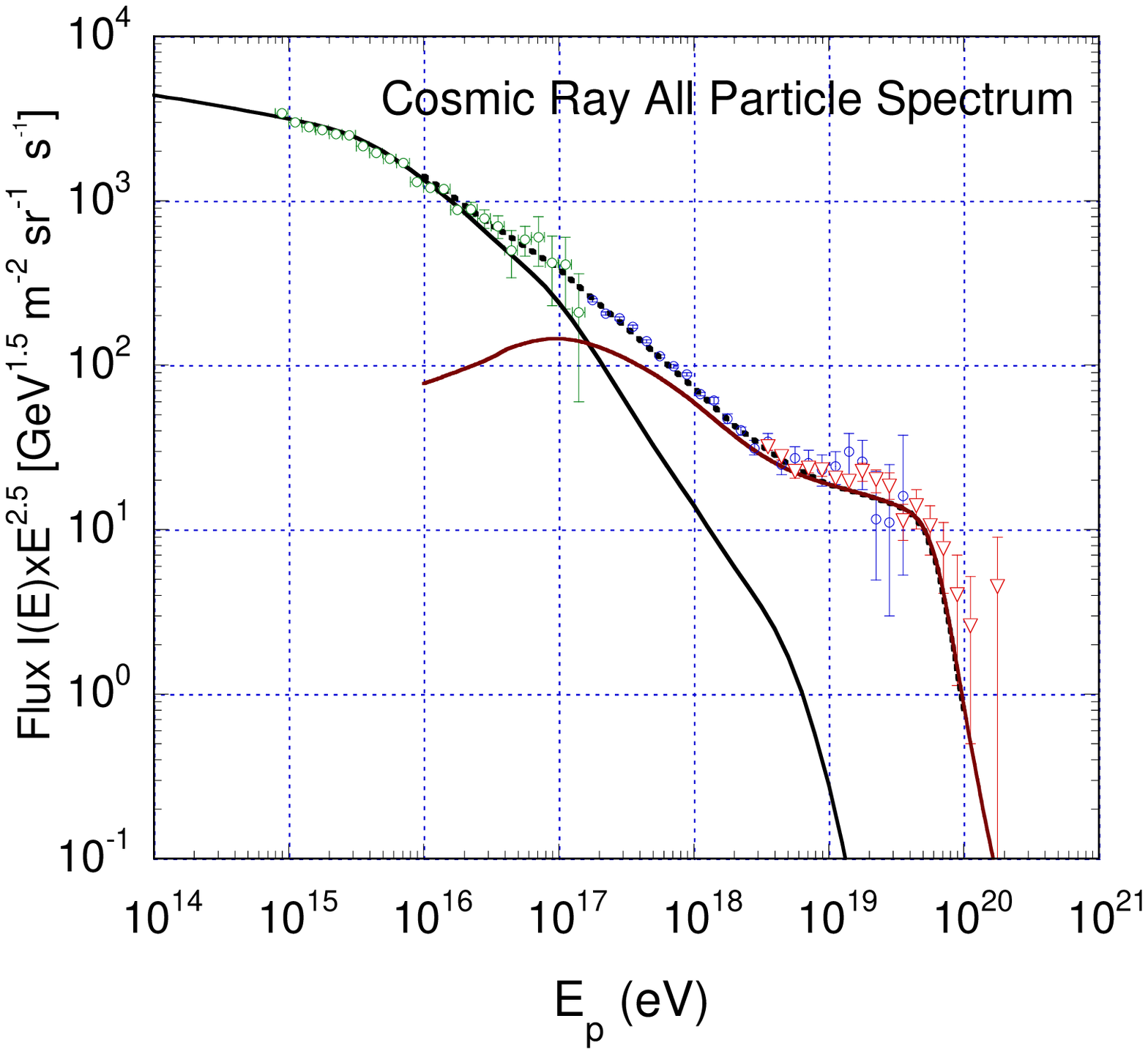}
}}
\caption{The all-particle CR data from KASCADE, High-Res I and II, 
in comparison with the model result for the all-particle spectrum
(dotted curve) from galactic (lower energy solid curve) and
extragalactic (higher energy solid curve) GRB sources of high-energy
cosmic rays.}
\label{fig:CRaps}
\end{figure}

In conclusion, we have investigated the hypothesis that GRBs are the
sources of HECRs.  Our model provides a unified source-type to fit all
the CR spectri, from some minimum CR energy produced in GRBs, which could
be somewhere $\lesssim 10^{14}$~eV, to a maximum energy $\gsim
10^{20}$~eV.  The total CR flux that we calculate is a superposition
of CRs originating from a past GRB in our Galaxy in whose HECR halo we
inhabit (with a gradual transition at lower energies to CRs
contributed from ordinary SN events with sub-relativistic blast
waves), and $\gtrsim 3\times 10^{17}$ eV CRs and UHECRs orginating in
extragalactic GRBs and GRBs at cosmological distances.  The HECR
all-particle spectrum and our model fit to this data are shown in
Fig. \ref{fig:CRaps}.

A GRB model for HECR production requires that GRB blast waves are
hadronically dominated by nearly 2 orders of magnitude. This could be
related to the injection process in relativistic shocks and the large
proton/electron mass ratio.  This has important consequences for 100
MeV -- GeV -- TeV $\gamma$-ray emission and high-energy ($\gtrsim 100$
TeV) neutrino detection. Allowed hadronic production in GRB blast
waves will make $\gamma$ rays of comparable fluence as the neutrino
fluence.  Insisting that the high-energy radiation fluence from a
photomeson cascade can only reach the level of the X/$\gamma$ emission
measured with BATSE (unless it is missed by detectors operating at
larger energies), we have shown that $\gtrsim 1$ neutrino could be
detected coincident with a GRB at the fluence level $\gtrsim 3\times
10^{-4}$ ergs cm$^{-1}$ by IceCube.

If GLAST shows that GRBs are much more fluent at $\gtrsim 100$ MeV
energies than at X/$\gamma$ energies, then in these GRBs one could
expect a few high-energy neutrinos.  Detection of even 1 or 2
neutrinos from GRBs with IceCube or a northern hemisphere neutrino
telescope would unambiguously demonstrate the high nonthermal baryon
load in GRBs.  High-energy neutrino detection, especially from GRBs
with bright 100 MeV -- GeV emission components, or from GRBs with
small Lorentz factors or large baryon-loading such as the X-ray bright
GRBs, would provide compelling support for this scenario for the
origin of the cosmic rays.

\vskip0.2in

We thank the referee for a helpful report. The work of S.D.W.~was
performed while he held a National Research Council Research
Associateship Award at the Naval Research Laboratory (Washington,
D.C). The work of C.D.D.\ is supported by the Office of Naval Research
and NASA {\it GLAST} science investigation grant DPR \#
S-15634-Y. A.A.\ acknowledges support and hospitality during visits to
the High Energy Space Environment Branch of the Naval Research
Laboratory.

\appendix

\section{UHECR Attenuation and Flux Calculation}

The Larmor radius of a UHE
proton is $\simeq 100~E_{20}/B_{\mu\rm{G}}$ kpc for a proton
of energy $E=10^{20}~E_{20}$ eV propagating through a
magnetic field of strength $B= B_{\mu\rm{G}}~\mu\rm{G}$.
We approximate the UHECRs as propagating
rectilinearly where their spectrum is modified by energy losses.
Here we describe the three attenuation processes included in our propagation 
code:  1) red-shifting of the proton momentum, 2) 
pion production from proton-CMBR scattering 
($p~\gamma~\rightarrow~\pi^+~n~{\rm{or}}~\pi^0~p$), 
and 3) electron-pair production
from proton scattering on the CMBR ($p~\gamma~\rightarrow~p~e^+~e^-$).

The expansion of the universe causes a red-shifting of the 
proton momentum 
\beq
-\frac{1}{E}\frac{dE_r(z)}{dt} = \frac{1}{(1+z)}\frac{dz}{dt}\; ,
\eeq
where for a $\Lambda$CDM cosmology, 
\beq
\frac{dz}{dt}= (1+z)H_0 \sqrt{\Omega_m(1+z)^{3}+\Omega_{\Lambda}} \;.
\label{eq:dtdz}
\eeq

Energy loss from proton scattering on CMBR is given by
\beq
-\frac{1}{E}\frac{dE_{p\gamma}(z)}{dt}= \int_{\epsilon_{th}/2\gamma_p}^{\infty}d\epsilon \;
\frac{n_{\gamma}(\epsilon,z)}{2\gamma_{p}^{2}\epsilon^2}
\int_{\epsilon_{th}}^{2\gamma_p\epsilon}d\epsilon_r\;
\sigma_{p\gamma}(\epsilon_r)K_{p\gamma}(\epsilon_r)\epsilon_r\;
\label{eq:eloss}
\eeq
\cite{bg88},
where $n_{\gamma}(\epsilon,z)$ is the black-body photon distribution of
the CMBR with red-shift dependent temperature 
$T(z)=2.72\rm{K}~(1+z)$,~
$\sigma_{p\gamma}(\epsilon_r)$ is the cross-section for the process, 
and $K_{p\gamma}(\epsilon_r)$ is the inelasticity.
The mean-free-path between scatterings is found
from eq.~(\ref{eq:eloss}) by setting $K_{p\gamma}(\epsilon_r)=1$ for
all $\epsilon_r.$  

We approximate photo-pion production following Ref.\ \cite{ad03}, 
modified to include three step cross-section and
inelasticity functions
\beq
\sigma_{p\gamma}(\epsilon_r)=
 \;\cases{ 
40 {\rm ~mb}\; ,& for $150 {\rm~ MeV} \leq \epsilon_r < 220 {\rm~ MeV}$~~,  \cr\cr 
300 {\rm ~mb}\; ,& for $220 {\rm~ MeV} \leq \epsilon_r < 450 {\rm~ MeV}$~~, \cr\cr 
110 {\rm ~mb}\; ,& for $450 {\rm~ MeV} \leq \epsilon_r$ ~~, \cr}
\; \; \;
\eeq
and 
\beq
K_{p\gamma}(\epsilon_r)=
 \;\cases{ 
0.2 \; ,& for $150 {\rm~ MeV} \leq \epsilon_r < 220 {\rm~ MeV}$ ~~,  \cr\cr 
0.2\; ,& for $220 {\rm~ MeV} \leq \epsilon_r < 450 {\rm~ MeV}$ ~~, \cr\cr 
0.5\; ,& for $450 {\rm~ MeV} \leq \epsilon_r$ ~~. \cr}
\; \; \;
\eeq
With increasing $\epsilon_r$, the three regions 
correspond to 1) direct pion production at threshold, 
2) the enhanced cross-section at the $\Delta$-resonance,
and 3) multi-pion production in the diffractive scattering regime where
photon coupling to the vector mesons $\rho_0$ and $\omega$ dominate.   
In comparison with the attenuation calculations of
M\"ucke et al.\ \cite{Muecke} and Stanev et al.\ \cite{sta00}, we agree to $\approx$ 2 (4)\% at
$E_p=10^{20}(10^{21})$~eV.

Our treatment of the photo-pair production 
losses follows that of Chodorowski et al.\ \cite{Chod}.  The greatest
photopair energy-loss rate is found at $E_p\simeq 3\times 10^{19}$ eV
when $z =0$, and pair-production dominates other energy losses over the range 
$2\times 10^{18}~\rm{eV}\lesssim E_p\lesssim 6\times 10^{19}~\rm{eV}.$

Our UHECR flux calculation follows the formalism
of Berezinsky and Grigor'eva \cite{bg88} for 
diffuse sources.  The flux from a volume element $dV(z)=R^3(z)r^2dr d\Omega$ 
observed at Earth is 
\beq
dJ(E)dE=\frac{n(z)dV(z)F(E^*,E_{max})dE^*}{(1+z)4\pi R_0^2 r^2}.
\eeq
where $n(z)$ is the comoving source density described in eq.~(\ref{eq:evol}) 
and Fig.\ \ref{fig:madau},
$F(E,E_{max})$ is the number of CRs produced
per unit energy per unit time for an injection spectrum with 
power-law index $p$ and exponential cutoff energy
$E_{\rm{max}},$ the Hubble length $R_0= 4286$ Mpc, and
$E^*$ is the proton energy at generation which is subsequently 
measured with an energy $E < E^*$.  
We use $R_0=R(z)(1+z)$ and $R(z)dr=cdt$ to obtain
\beq
J(E)=\frac{1}{4\pi}~F(E,E_{max})\int_{0}^{z_{max}}dz~
\frac{dt}{dz}~n(z)~\left(\frac{E^*}{E}\right)^{-p}
~\left|\frac{dE^*(E,z)}{dE}\right|\;,
\label{eq:flux}
\eeq
where the term $|dE^*(E,z)/dE|$ transforms the energy 
interval from propagation losses and we take $z_{max}=4.0$.
Results of calculating eq.~(\ref{eq:flux}) for several different spectra and evolution
models are found in Figs.\ \ref{fig:cr2}, \ref{fig:cr3}, \ref{fig:cr4}, and \ref{fig:crAGASA}.

\end{document}